
\input harvmac

\input epsf
\epsfverbosetrue
\def\epsfsize#1#2{\hsize}

\def\v#1{{\bf #1}}
\def\B{\bullet}
\def\W{\circ}

\bigskip

 \font\dynkfont=cmsy10 scaled\magstep4    \skewchar\dynkfont='60
\def\dynk{\textfont2=\dynkfont}
\def\hr#1,#2;{\dimen0=.4pt\advance\dimen0by-#2pt
              \vrule width#1pt height#2pt depth\dimen0}
\def\vr#1,#2;{\vrule height#1pt depth#2pt}
\def\blb#1#2#3#4#5
            {\hbox{\ifnum#2=0\hskip11.5pt
                   \else\ifnum#2=1\hr13,5;\hskip-1.5pt
                   \else\ifnum#2=2\hr13.5,6.5;\hskip-13.5pt
                                  \hr13.5,3.5;\hskip-2pt
                   \else\ifnum#2=3\hr13.7,8;\hskip-13.7pt
                                  \hr13,5;\hskip-13pt
                                  \hr13.7,2;\hskip-2.2pt\fi\fi\fi\fi
                   $#1$
                   \ifnum#4=0
                   \else\ifnum#4=1\hskip-9.2pt\vr22,-9;\hskip8.8pt
                   \else\ifnum#4=2\hskip-10.9pt\vr22,-8.75;\hskip3pt
                                  \vr22,-8.75;\hskip7.1pt
                   \else\ifnum#4=3\hskip-12.6pt\vr22,-8.5;\hskip3pt
                                  \vr22,-9;\hskip3pt
                                  \vr22,-8.5;\hskip5.4pt\fi\fi\fi\fi
                   \ifnum#5=0
                   \else\ifnum#5=1\hskip-9.2pt\vr1,12;\hskip8.8pt
                   \else\ifnum#5=2\hskip-10.9pt\vr1.25,12;\hskip3pt
                                  \vr1.25,12;\hskip7.1pt
                   \else\ifnum#5=3\hskip-12.6pt\vr1.5,12;\hskip3pt
                                  \vr1,12;\hskip3pt
                                  \vr1.5,12;\hskip5.4pt\fi\fi\fi\fi
                   \ifnum#3=0\hskip8pt
                   \else\ifnum#3=1\hskip-5pt\hr13,5;
                   \else\ifnum#3=2\hskip-5.5pt\hr13.5,6.5;
                                  \hskip-13.5pt\hr13.5,3.5;
                   \else\ifnum#3=3\hskip-5.7pt\hr13.7,8;
                                  \hskip-13pt\hr13,5;
                                  \hskip-13.7pt\hr13.7,2;\fi\fi\fi\fi }}
\def\blob#1#2#3#4#5#6#7{\hbox
{$\displaystyle\mathop{\blb#1#2#3#4#5 }_{#6}\sp{#7}$}}
\def\up#1#2{\dimen1=33pt\multiply\dimen1by#1\hbox{\raise\dimen1\rlap{#2}}}
\def\uph#1#2{\dimen1=17.5pt\multiply\dimen1by#1\hbox{\raise\dimen1\rlap{#2}}}
\def\dn#1#2{\dimen1=33pt\multiply\dimen1by#1\hbox{\lower\dimen1\rlap{#2}}}
\def\dnh#1#2{\dimen1=17.5pt\multiply\dimen1by#1\hbox{\lower\dimen1\rlap{#2}}}

\def\rlbl#1{\kern-8pt\raise3pt\hbox{$\scriptstyle #1$}}
\def\llbl#1{\raise3pt\llap{\hbox{$\scriptstyle #1$\kern-8pt}}}
\def\elbl#1{\kern3pt\lower4.5pt\hbox{$\scriptstyle #1$}}
\def\lelbl#1{\rlap{\hbox{\kern-9pt\raise2.5pt\hbox{{$\scriptstyle #1$}}}}}

\def\whtd#1#2#3#4#5{\blob\circ#1#2#3#4{#5}{}}
\def\blkd#1#2#3#4#5{\blob\bullet#1#2#3#4{#5}{}}
\def\whtu#1#2#3#4#5{\blob\circ#1#2#3#4{}{#5}}
\def\blku#1#2#3#4#5{\blob\bullet#1#2#3#4{}{#5}}
\def\whtr#1#2#3#4#5{\blob\circ#1#2#3#4{}{}\rlbl{#5}}

\def\rwng{\hbox{$\vbox{\offinterlineskip{
  \hbox{\phantom{}\kern6pt{$\circ$}}\kern-2.5pt\hbox{$\Biggr/$}\kern-0.5pt
  \hbox{\phantom{}\kern-5pt$\circ$}\kern-3.0pt\hbox{$\Biggr\backslash$}
  \kern-1.5pt\hbox{\phantom{}\kern6pt{$\circ$}} }}$}}

\def\lwng{\hbox{$\vbox{\offinterlineskip{ \hbox{$\circ$}
  \kern-3.0pt\hbox{\phantom{}\kern6.0pt{$\Biggr\backslash$}}
  \kern-0.5pt\hbox{\phantom{}\kern11pt{$\circ$}}\kern-3.5pt
  \hbox{\phantom{}\kern5.0pt {$\Biggr/$}}\kern-1.0pt\hbox{$\circ$} }}$}}

 \def\drwng#1#2#3{\hbox{$\vcenter{ \offinterlineskip{
  \hbox{\phantom{}\kern6pt{$\circ^{\elbl{#3}}$}}
  \kern-2.5pt\hbox{$\Biggr/$}\kern-0.5pt
  \hbox{\phantom{}\kern-5pt$\circ^{ \elbl{#1}}$}
  \kern-3.0pt\hbox{$\Biggr\backslash$}
  \kern-1.5pt\hbox{\phantom{}\kern6pt{$\circ^{\elbl{#2}}$}}  } }$}}

\def\dlwng#1#2#3{\hbox{$\vcenter{\offinterlineskip{ \hbox{$\lelbl{#1}\circ$}
  \kern-3.0pt\hbox{\phantom{}\kern6.0pt{$\Biggr\backslash$}}
  \kern-0.5pt\hbox{\phantom{}\kern11pt{$\lelbl{#2}\circ$}}\kern-3.5pt
  \hbox{\phantom{}\kern5.0pt {$\Biggr/$}}\kern-1.0pt\hbox{$\lelbl{#3}
  \circ$}}}$} }

\def\rde#1#2#3{\hbox{\phantom{}\kern-4pt\hbox{$\vcenter{\offinterlineskip
\hbox{
               \raise 4.5pt\hbox{\vrule height0.4pt width13pt depth0pt}
                \kern-1pt\vbox{ \hbox{\drwng{#1}{#2}{#3}}} }}$  }}  }

\def\lde#1#2#3{\hbox{$\vcenter{\offinterlineskip  \hbox{
               \dlwng{#1}{#2}{#3}\kern-4.2pt\lower0.4pt\hbox{$\vcenter{\hrule
width13pt}$}
               \kern-8pt\phantom{}   }}  $}}

\def\rwngb{\hbox{$\vbox{\offinterlineskip{
  \hbox{\phantom{}\kern6pt{$\bullet$}}\kern-2.5pt\hbox{$\Biggr/$}\kern-0.5pt
  \hbox{\phantom{}\kern-5pt$\bullet$}\kern-3.0pt\hbox{$\Biggr\backslash$}
  \kern-1.5pt\hbox{\phantom{}\kern6pt{$\bullet$}} }}$}}

\def\lwngb{\hbox{$\vbox{\offinterlineskip{ \hbox{$\bullet$}
  \kern-3.0pt\hbox{\phantom{}\kern6.0pt{$\Biggr\backslash$}}
  \kern-0.5pt\hbox{\phantom{}\kern11pt{$\bullet$}}\kern-3.5pt
  \hbox{\phantom{}\kern5.0pt {$\Biggr/$}}\kern-1.0pt\hbox{$\bullet$} }}$}}

\def\dbrwng#1#2#3{\hbox{$\vcenter{ \offinterlineskip{
  \hbox{\phantom{}\kern6pt{$\bullet^{\elbl{#3}}$}}
  \kern-2.5pt\hbox{$\Biggr/$}\kern-0.5pt
  \hbox{\phantom{}\kern-5pt$\bullet^{ \elbl{#1}}$}
  \kern-3.0pt\hbox{$\Biggr\backslash$}
  \kern-1.5pt\hbox{\phantom{}\kern6pt{$\bullet^{\elbl{#2}}$}}  } }$}}

\def\dblwng#1#2#3{\hbox{$\vcenter{\offinterlineskip{ \hbox{$\lelbl{#1}\bullet$}
  \kern-3.0pt\hbox{\phantom{}\kern6.0pt{$\Biggr\backslash$}}
  \kern-0.5pt\hbox{\phantom{}\kern11pt{$\lelbl{#2}\bullet$}}\kern-3.5pt
  \hbox{\phantom{}\kern5.0pt
{$\Biggr/$}}\kern-1.0pt\hbox{$\lelbl{#3}\bullet$}}}$} }

\def\rbde#1#2#3{\hbox{\phantom{}\kern-4pt\hbox{$\vcenter{\offinterlineskip
\hbox{
               \raise 4.5pt\hbox{\vrule height0.4pt width13pt depth0pt}
                \kern-1pt\vbox{ \hbox{\dbrwng{#1}{#2}{#3}}} }}$  }}  }

\def\lbde#1#2#3{\hbox{$\vcenter{\offinterlineskip  \hbox{
               \dblwng{#1}{#2}{#3}\kern-4.2pt\lower0.4pt\hbox{$\vcenter{\hrule
width13pt}$}
               \kern-8pt\phantom{}   }}  $}}

\def\ddgu#1.#2.{\dynk  \whtu0300{#1}\blku3000{#2}}
\def\ddgd#1.#2.{\dynk  \whtd0300{#1}\blkd3000{#2}}

\def\eddgiu#1.#2.#3.{\dynk \whtu0100{#1}\whtu1300{#2}\blku3000{#3}}
\def\eddgid#1.#2.#3.{\dynk \whtd0100{#1}\whtd1300{#2}\blkd3000{#3}}
\def\eddgiiu#1.#2.#3.{\dynk  \whtu0300{#1}\blku3100{#2}\blku1000{#3}}
\def\eddgiid#1.#2.#3.{\dynk  \whtd0300{#1}\blkd3100{#2}\blkd1000{#3}}

\def\ddfu#1.#2.#3.#4.{\dynk
\whtu0100{#1}\whtu1200{#2}\blku2100{#3}\blku1000{#4}}
\def\ddfd#1.#2.#3.#4.{\dynk
\whtd0100{#1}\whtd1200{#2}\blkd2100{#3}\blkd1000{#4}}

\def\eddfiu#1.#2.#3.#4.#5.{\dynk
\whtu0100{#1}\whtu1100{#2}\whtu1200{#3}\blku2100{#4}\blku1000{#5}}
\def\eddfid#1.#2.#3.#4.#5.{\dynk
\whtd0100{#1}\whtd1100{#2}\whtd1200{#3}\blkd2100{#4}\blkd1000{#5}}
\def\eddfiiu#1.#2.#3.#4.#5.{\dynk
\whtu0100{#1}\whtu1200{#2}\blku2100{#3}\blku1100{#4}\blku1000{#5}}
\def\eddfiid#1.#2.#3.#4.#5.{\dynk
\whtd0100{#1}\whtd1200{#2}\blkd2100{#3}\blkd1100{#4}\blkd1000{#5}}

\def\ddanu#1.#2.#3.#4.#5.{\dynk \whtu0100{#1}\whtu1100{#2}\whtu1100{#3}\cdots
                           \whtu1100{#4}\whtu1000{#5}}
\def\ddand#1.#2.#3.#4.#5.{\dynk \whtd0100{#1}\whtd1100{#2}\whtd1100{#3}\cdots
                           \whtd1100{#4}\whtd1000{#5}}

\def\eddanu#1.#2.#3.#4.#5.{\dynk \whtu0100{#1}\whtu1100{#2}%
                           \up1{\whtr0000{#3}}\cdots\whtu1100{#4}\whtu1000{#5}}
\def\eddand#1.#2.#3.#4.#5.{\dynk \whtd0100{#1}\whtd1100{#2}%
                           \up1{\whtr0000{#3}}\cdots\whtd1100{#4}\whtd1000{#5}}

\def\eddanid#1.#2.#3.#4.#5.{\dynk \whtd0200{#1}\whtd2100{#2}%
                           \whtd1100{#3}\cdots\whtd1200{#4}\blkd2000{#5}}
\def\eddaniu#1.#2.#3.#4.#5.{\dynk \whtu0200{#1}\whtu2100{#2}%
                           \whtu1100{#3}\cdots\whtu1200{#4}\blku2000{#5}}

\def\eddaniid#1.#2.#3.#4.#5.#6.{\hbox{$\vcenter{\hbox
         {\dynk\hbox{$ \lbde{#1}{#2}{#3}\blkd1100{#4}\cdots%
          \blkd1200{#5}\whtd2000{#6} $}} }$}}
\def\eddaniiu#1.#2.#3.#4.#5.#6.{\hbox{$\vcenter{\hbox
         {\dynk\hbox{$ \lbde{#1}{#2}{#3}\blku1100{#4}\cdots%
          \blku1200{#5}\whtu2000{#6} $}} }$}}

\def\ddbnu#1.#2.#3.#4.#5.{\dynk \whtu0100{#1}\whtu1100{#2}\whtu1100{#3}\cdots
                           \whtu1200{#4}\blku2000{#5}}
\def\ddbnd#1.#2.#3.#4.#5.{\dynk \whtd0100{#1}\whtd1100{#2}\whtd1100{#3}\cdots
                           \whtd1200{#4}\blkd2000{#5}}

\def\eddbnu#1.#2.#3.#4.#5.#6.{\dynk \lde{#1}{#2}{#3}\whtu1100{#4}\cdots
                           \whtu1200{#5}\blku2000{#6}}
\def\eddbnd#1.#2.#3.#4.#5.#6.{\dynk \lde{#1}{#2}{#3}\whtd1100{#4}\cdots
                           \whtd1200{#5}\blkd2000{#6}}

\def\ddcnu#1.#2.#3.#4.#5.{\dynk \blku0100{#1}\blku1100{#2}\blku1100{#3}\cdots
                           \blku1200{#4}\whtu2000{#5}}
\def\ddcnd#1.#2.#3.#4.#5.{\dynk \blkd0100{#1}\blkd1100{#2}\blkd1100{#3}\cdots
                           \blkd1200{#4}\whtd2000{#5}}

\def\eddcnu#1.#2.#3.#4.#5.#6.{\dynk \whtu0200{#1}\blku2100{#2}\blku1100{#3}
       \blku1100{#4}\cdots
                           \blku1200{#5}\whtu2000{#6}}

\def\eddcnd#1.#2.#3.#4.#5.{\dynk \whtd0200{#1}\blkd2100{#2}\blkd1100{#3}
       \cdots \blkd1200{#4}\whtd2000{#5}}

\def\dddnu#1.#2.#3.#4.#5.#6.{\hbox{$\vcenter{\hbox
         {\dynk\hbox{$ \whtu0100{#1}\whtu1100{#2}\cdots%
          \whtu1100{#3}\rde{#4}{#5}{#6} $}}  }$}}
\def\dddnd#1.#2.#3.#4.#5.#6.{\hbox{$\vcenter{\hbox
         {\dynk\hbox{$ \whtd0100{#1}\whtd1100{#2}\cdots%
          \whtd1100{#3}\rde{#4}{#5}{#6} $}} }$}}
\def\dddiv#1.#2.#3.#4.{\hbox{$\vcenter{\hbox
         {\dynk\hbox{$ \whtu0100{#1}\rde{#2}{#3}{#4}
              $}}  }$}}

\def\edddnu#1.#2.#3.#4.#5.#6.#7.#8.{\hbox{$\vcenter{\hbox
         {\dynk\hbox{$ \lde{#1}{#2}{#3}\whtu1100{#4}\cdots%
          \whtu1100{#5}\rde{#6}{#7}{#8} $}}  }$}}
\def\edddnd#1.#2.#3.#4.#5.#6.#7.#8.{\hbox{$\vcenter{\hbox
         {\dynk\hbox{$ \lde{#1}{#2}{#3}\whtd1100{#4}\cdots%
          \whtd1100{#5}\rde{#6}{#7}{#8} $}} }$}}

\def\edddniid#1.#2.#3.#4.#5.{\hbox{$\vcenter{\hbox
         {\dynk\hbox{$ \blkd0200{#1}\whtd2100{#2}\whtd1100{#3}\cdots%
          \whtd1200{#4}\blkd2000{#5} $}} }$}}
\def\edddniiu#1.#2.#3.#4.#5.{\hbox{$\vcenter{\hbox
         {\dynk\hbox{$ \blku0200{#1}\whtu2100{#2}\whtu1100{#3}\cdots%
          \whtu1200{#4}\blku2000{#5} $}} }$}}

\def\ddei#1.#2.#3.#4.#5.#6.{\hbox{$\vcenter{\hbox
       {\dynk \whtd0100{#1}\whtd1100{#3}%
       \up1{\whtr0001{#2}}\whtd1110{#4}\whtd1100{#5}\whtd1000{#6}} }$}}

\def\eddei#1.#2.#3.#4.#5.#6.#7.{\hbox{$\vcenter{\hbox
       {\dynk \whtd0100{#1}\whtd1100{#3}%
       \up1{\whtr0011{#2}}\up2{\whtr0001{#7}}\whtd1110{#4}\whtd1100{#5}%
       \whtd1000{#6}} }$}}


\def\ddeii#1.#2.#3.#4.#5.#6.#7.{\hbox{$\vcenter{\hbox
       {\dynk \whtd0100{#1}\whtd1100{#3}%
       \up1{\whtr0001{#2}}\whtd1110{#4}\whtd1100{#5}\whtd1100{#6}%
       \whtd1000{#7}} }$}}

\def\eddeii#1.#2.#3.#4.#5.#6.#7.#8.{\hbox{$\vcenter{\hbox
       {\dynk \whtd0100{#8}\whtd1100{#1}\whtd1100{#3}%
       \up1{\whtr0001{#2}}\whtd1110{#4}\whtd1100{#5}\whtd1100{#6}%
       \whtd1000{#7}} }$}}

\def\ddeiii#1.#2.#3.#4.#5.#6.#7.#8.{\hbox{$\vcenter{\hbox
       {\dynk \whtd0100{#1}\whtd1100{#3}%
       \up1{\whtr0001{#2}}\whtd1110{#4}\whtd1100{#5}\whtd1100{#6}%
       \whtd1100{#7}\whtd1000{#8}} }$}}

\def\eddeiii#1.#2.#3.#4.#5.#6.#7.#8.#9.{\hbox{$\vcenter{\hbox
       {\dynk \whtd0100{#1}\whtd1100{#3}%
       \up1{\whtr0001{#2}}\whtd1110{#4}\whtd1100{#5}\whtd1100{#6}%
       \whtd1100{#7}\whtd1100{#8}\whtd1000{#9}} }$}}


%
%
\def\RF#1#2{\if*#1\ref#1{#2.}\else#1\fi}
\def\NRF#1#2{\if*#1\nref#1{#2.}\fi}
\def\refdef#1#2#3{\def#1{*}\def#2{#3}}
%
%
\def \ts{\thinspace}

\def \AJM{{\it Am.\ts J.\ts Math.\ts}}
\def \CMP{{\it Commun.\ts Math.\ts Phys.\ts }}
\def \NP{{\it Nucl.\ts Phys.\ts }}
\def \PL{{\it Phys.\ts Lett.\ts }}
\def \PR{{\it Phys.\ts Rev.\ts }}
\def \Tahoe{Proceedings of the XVIII
 International Conference on Differential Geometric Methods in Theoretical
 Physics: Physics and Geometry, Lake Tahoe, USA 2-8 July 1989}
\def \Tahoe{Proceedings of the NATO
 Conference on Differential Geometric Methods in Theoretical
 Physics, Lake Tahoe, USA 2-8 July 1989 (Plenum 1990)}
\def \Zm{Zamolodchikov}
\def \AZm{A.\ts B.\ts \Zm}
\def \AlZm{Al.\ts B.\ts \Zm}
\def\dur{H.\ts W.\ts Braden, E.\ts Corrigan, P.\ts E.\ts Dorey \ and R.\ts
Sasaki}
%
%
\refdef\rAFZa\AFZa{A.\ts E.\ts Arinshtein, V.\ts A.\ts Fateev and
 \AZm, \PL {\bf B87} (1979) 389}

\refdef\rACFGZ\ACFGZ{H. Aratyn, C.P. Constantinidis, L.A. Ferreira, J.F. Gomes
and
A.H. Zimerman, \NP {\bf B406} (1993) 727}

\refdef\rBa\Ba{H.W. Braden, {\it J. Phys.} {\bf A25} (1992) L15}

\refdef\rBc\Bc{N.\ts Bourbaki, {\it Groupes et alg\`ebres de Lie} {\bf
 IV, V, VI,} (Hermann, Paris 1968)}
\refdef\rBd\Bd{L. Bonora, {\it Int. J Mod. Phys.} {\bf B6} (1992) 2015}

\refdef\rBBa\BBa{O. Babelon and L. Bonora, \PL {\bf B267} (1991) 71}

\refdef\rBCDSa\BCDSa{\dur, \PL {\bf B227} (1989) 411}

\refdef\rBCDSb\BCDSb{\dur, \Tahoe}

\refdef\rBCDSc\BCDSc{\dur, \NP {\bf B338} (1990) 689}

\refdef\rBCDSe\BCDSe{\dur, \NP {\bf B356} (1991) 469}

\refdef\rBCKKSa\BCKKSa{H.W. Braden, H.S. Cho, J.D. Kim, I.G. Koh and
R. Sasaki, {\it Prog. Theor. Phys. } {\bf 88} (1992) 1205}

\refdef\rBCGTa\BCGTa{E. Braaten, T. Curtright, G. Ghandour and C. Thorn, {\it
 Phys. Lett}. {\bf B125} (1983) 301}

\refdef\rBLa\BLa{D. Bernard and A. LeClair, \CMP {\bf 142} (1991) 99 }

\refdef\rBLb\BLb{D. Bernard and A. LeClair, \NP {\bf B399} (1993) 709}

\refdef\rBSa\BSa{H.W. Braden and R. Sasaki, \PL {\bf B255} (1991) 343}

\refdef\rBSb\BSb{H.W. Braden and R. Sasaki, \NP {\bf B379} (1992) 377}

\refdef\rCa\Ca{P. Christe,  \Tahoe}

\refdef\rCi\Ci{\Cg\semi\Ch}

\refdef\rCDSa\CDSa{E. Corrigan, P.E. Dorey and R. Sasaki,
 \NP {\bf B408} (1993) 579}

\refdef\rCDRSa\CDRSa{E. Corrigan, P.E. Dorey, R. Rietdijk and R. Sasaki,
\PL {\bf B333} (1994) 83}

\refdef\rCDRa\CDRa{E. Corrigan, P.E. Dorey and R. Rietdijk, {\it
Aspects of affine Toda field theory on a half line}, Durham preprint DTP-94/29,
hep-th/9407148}

\refdef\rCKKa\CKKa{H.S. Cho, I.G. Koh and J.D. Kim,
\PR {\bf D47} (1993) 2625}

\refdef\rCMa\CMa{P.\ts Christe and G.\ts Mussardo, {\it Nucl. Phys}.
 {\bf B330} (1990) 465}

\refdef\rCMb\CMb{P.\ts Christe and G.\ts Mussardo,
 {\it Int.~J.~Mod.~Phys.}~{\bf A5} (1990) 4581}

\refdef\rCTa\CTa{S.\ts Coleman and H.\ts Thun, \CMP {\bf 61}
 (1978) 31}

\refdef\rCZa\CZa{Z. Zhu and D.G. Caldi, {\it Multi-soliton solutions of affine
Toda models},
SUNY, hep-th/9307175}

\refdef\rDc\Dc{P.\ts E.\ts Dorey ,  \NP {\bf B358} (1991) 654}

\refdef\rDd\Dd{P.\ts E.\ts Dorey , \NP {\bf B374} (1992) 741}

\refdef\rDe\De{P.\ts E.\ts Dorey , \PL {\bf B312} (1993) 291}

\refdef\rDf\Df{P.\ts E.\ts Dorey , In the proceedings of the conference
``Integrable Quantum Field Theories'',
Como, Italy, 13-19 September 1992 (Plenum, 1993)}

\refdef\rDDa\DDa{C.\ts Destri and H.\ts J.\ts de Vega, {\it Phys. Lett.}
 {\bf B233} (1989) 336}

\refdef\rDDFa\DDFa{C. Destri, H.J. de Vega and V.A. Fateev, \PL {\bf
256B} (1991) 173}

\refdef\rDMa\DMa{G. Delfino and G. Mussardo, \PL {B324} (1994) 40}
\refdef\rDGPZa\DGPZa{G.W. Delius, M.T. Grisaru, S. Penati and D. Zanon,
\PL {\bf 256B} (1991) 164}
\refdef\rDGZa\DGZa{G.W. Delius, M.T. Grisaru and D. Zanon, \NP {\bf B382}
(1992) 365}

\refdef\rDGa\DGa{G.W. Delius and M.T. Grisaru, {\it Toda soliton mass
corrections
and the particle-soliton duality conjecture}, King's College preprint,
hep-th/9411176}

\refdef\rEYa\EYa{T. Eguchi and S-K Yang, {\it Phys. Lett.} {\bf  B224} (1989)
 373}

\refdef\rFb\Fb{M.\ts D.\ts Freeman, \PL {\bf B261} (1991) 57}

\refdef\rFMsa\FMSa{A. Fring, G. Mussardo and P, Simonetti, \PL {B307} (1993)
83}

\refdef\rFKc\FKc{A.\ts Fring and R.\ts K\"oberle, \NP {\bf B421} (1994) 159}

\refdef\rFKd\FKd{A.\ts Fring and R.\ts K\"oberle, \NP {\bf B419} (1994) 647}
\refdef\rGd\Gd{S.\ts Ghoshal, {\it Int. J. Mod. Phys.} {\bf A9} (1994) 4801}

\refdef\rGZa\GZa{S.\ts Ghoshal and \AZm ,{\it Int. J. Mod. Phys.}
{\bf A9} (1994) 3841}

\refdef\rFKMa\FKMa{P. G. O. Freund, T. Klassen and E. Melzer, {\it Phys. Lett.}
 {\bf B229} (1989) 243}

\refdef\rFOa\FOa{A. Fring and D.I. Olive,
\NP {\bf B379} (1992) 429}

\refdef\rFLOa\FLOa{A. Fring, H.C. Liao and D.I. Olive,
\PL {\bf 266B} (1991) 82}

\refdef\rFZa\FZa{V.\ts A.\ts Fateev and \AZm, {\it Int. J. Mod. Phys.} {\bf A5}
 (1990) 1025}

\refdef\rGNa\GNa{J-L. Gervais and A. Neveu, {\it Nucl. Phys.} {\bf B224} (1983)
 329}

\refdef\rHa\Ha{T.\ts J.\ts Hollowood, \NP {\bf 384} (1992) 523}

\refdef\rHb\Hb{T.J. Hollowood, {\it Int. J. Mod. Phys.} {\bf A8} (1993) 947}
\refdef\rHf\Hf{T.J. Hollowood, \PL {\bf B300} (1993) 73}

\refdef\rHd\Hd{J. E. Humphreys, {\it Reflection Groups and Coxeter Groups},
(Cambridge University Press 1990)}

\refdef\rHMa\HMa{T.\ts J.\ts Hollowood and P.\ts Mansfield, \PL {\bf B226}
 (1989) 73}

\refdef\rKa\Ka{M.\ts Karowski, \NP {\bf B153} (1979) 244}

\refdef\rKb\Kb{B.\ts Kostant, \AJM {\bf 81} (1959) 973}

\refdef\rKSRJa\KSRJa{\KSRa\semi\Jb}

\refdef\rKc\Kc{V.\ts Kac, {\it Infinite Dimensional Lie Algebras} (Birkhauser
Verlag 1983)}

\refdef\rKd\Kd{A. Koubek, \NP {\bf B428} (1994) 655}

\refdef\rKMa\KMa{T.\ts R.\ts Klassen and E.\ts Melzer, {\it Nucl. Phys.}
 {\bf B338} (1990) 485}

\refdef\rKWa\KWa{H. Kausch and G.M.T. Watts, \NP {\bf B386} (1992) 166}

\refdef\rLSa\LSa{A. N. Leznov and M. V. Saveliev,
{\it Group-theoretical methods for the
 integration of nonlinear dynamical systems} (Birkh\"auser Verlag 1992)}
\refdef\rMMa\MMa{N.J. Mackay and W.A. McGhee, {\it Inter. J. Mod. Phys.} {\bf
A8}
(1993) 2791}

\refdef\rMa\Ma{P. Mansfield, {\it Nucl. Phys.} {\bf B222} (1983) 419}

\refdef\rMb\Mb{G. Mussardo,  \Tahoe}

\refdef\rMc\Mc{W.A. McGhee, {\it Int. J. Mod. Phys.} {\bf A9} (1994) 2666 }

\refdef\rMCa\MCa{\Ca\semi\Mb\semi\CMb}

\refdef\rMd\Md{N.\ts J.\ts MacKay, \NP {\bf B356} (1991) 729}

\refdef\rMg\Mg{W.A. McGhee, Durham PhD Thesis, 1994}

\refdef\rMOPa\MOPa{A. V. Mikhailov, M. A. Olshanetsky and A. M. Perelomov, {\it
 Comm. Math. Phys.} {\bf 79} (1981) 473}

\refdef\rMWa\MWa{N.J. Mackay and G.M.T. Watts,
{\it Quantum mass corrections for affine Toda solitons}, DAMTP-94-36,
hep-th/9411169}

\refdef\rOTa\OTa{D. I. Olive and N. Turok, {\it Nucl. Phys.} {\bf B215} (1983)
 470}

\refdef\rOTb\OTb{D. I. Olive and N. Turok, {\it Nucl. Phys.} {\bf B265}
(1986) 469}

\refdef\rOTd\OTd{D. Olive and N. Turok, \NP {\bf B257} (1985) 277}

\refdef\rOTUb\OTUb{D.I. Olive, N. Turok and J.W.R. Underwood,
\NP {\bf B409} (1993) 509}

\refdef\rOTUa\OTUa{D.I. Olive, N. Turok and J.W.R. Underwood,
    \NP {\bf B401} (1993) 663}

\refdef\rSf\Sf{F.A. Smirnov, in {Introduction to Quantum Groups and
Integrable Massive Models of Quantum Field Theory}, Nankai Lectures in
Mathematical Physics, Mo-Lin Ge and Bao-Heng Zhao (eds), (World Scientific
1990)}

\refdef\rSg\Sg{F.A. Smirnov, {\it Form factors in completely integrable models
of
quantum field theory}, (World Scientific 1992)}

\refdef\rSk\Sk{R.\ts Sasaki, in Proceedings of the
conference ``Interface between physics and mathematics",
Hangzhou, China, 6-17 September 1993, (World Scientific 1994)}

\refdef\rSh\Sh{F.A. Smirnov, {\it Int. J. Mod. Phys}. {\bf A9} (1994) 5121}

\refdef\rSZb\SZb{R. Sasaki and F.P. Zen, {\it Int. J. Mod. Phys. }
{\bf A8} (1993) 115}

\refdef\rWa\Wa{G. Wilson, {\it Ergod. Th. and Dynam. Sys.} {\bf 1} (1981) 361}

\refdef\rWb\Wb{G.\ts Watts, \PL {\bf B245} (1990) 65}
\refdef\rWd\Wd{G.\ts Watts, \PL {\bf B338} (1994) 40}

\refdef\rWKa\WKa{K. Wilson and J. Kogut, {\it Phys. Rep.} {\bf 12C} (1974) 75}

\refdef\rWWa\WWa{G.M.T. Watts,  R. A.  Weston,
               \PL {\bf B289} (1992) 61}

\refdef\rZb\Zb{\AZm, {\it Int. J. Mod. Phys.} {\bf A4} (1989) 4235}

\refdef\rZd\Zd{\AZm, {\it Int. J. Mod. Phys.} {\bf A3} (1988) 743}

\refdef\rZz\Zz{\Za\semi\Zb}

\refdef\rZg\Zg{\AlZm, ``Thermodynamic Bethe Ansatz in Relativistic Models.
 Scaling 3-state Potts and Lee-Yang Models", Moscow ITEP preprint (1989)}

\refdef\rZZa\ZZa{\AZm\  and \AlZm, {\it Ann. Phys.}
 {\bf 120} (1979) 253}

\hfill DTP-94/55

\hfill hepth/9412213
\bigskip
\noindent {\bf Recent developments in affine Toda quantum field theory}
\vskip 3pc
\noindent {E. Corrigan}
\bigskip
\noindent{\obeylines
Department of Mathematical Sciences
University of Durham
South Road
Durham DH1 3LE
UK}


\vskip 3pc
\noindent Invited lectures at the CRM-CAP Summer School
\lq Particles and Fields 94'

\noindent August 16-24, 1994

\noindent Banff, Alberta, Canada.
\vfill
\noindent December 1994
\eject

\newsec{Introduction}

It is not intended in these four talks to give a detailed review of all
the recent activities in the area of Toda field theory, but
rather to highlight some of the  interesting developments,
and to point out some of the currently outstanding problems.
The list of references is by no means exhaustive.

Affine Toda field
theory\NRF\rAFZa\AFZa\NRF\rMOPa{\MOPa\semi\Wa\semi\OTa} \refs{\rAFZa,\rMOPa}
is a theory of $r$ scalar fields in
two-dimensional Minkowski space-time, where $r$ is the rank of a compact
semi-simple Lie algebra $g$. The classical field theory is determined by the
lagrangian density
\eqn\ltoda{{\cal L}={1\over 2} \partial_\mu\phi^a\partial^\mu\phi^a-V(\phi )}
where
\eqn\vtoda{V(\phi )={m^2\over\beta^2}\sum_0^rn_ie^{\beta\alpha_i\cdot\phi}.}
In \vtoda , $m$ and $\beta$ are real (classically unimportant) constants,
$\alpha_i\ i=1,\dots ,r$ are the simple roots of the Lie algebra $g$,
and $\alpha_0=-\sum_1^rn_i\alpha_i$ is a
linear combination of the simple roots; it corresponds to the extra spot
on an extended Dynkin diagram for $g$, at least in so far as
representing its inner products with the simple roots is concerned.
For notational reasons, $n_0=1$ in \vtoda , but the other integers $n_i$ are
characteristic for each
type of theory. They are tabulated in many places, for example in Kac' book
\NRF\rKc\Kc\refs{\rKc}.
The quantity $h=\sum_0^r n_i$ is called the Coxeter number.
For most purposes, in the present context
$\alpha_0$ will not represent a simple root of the affine algebra
$\widehat g$.

If the term $i=0$ is omitted from
\vtoda\ in
the lagrangian \ltoda , then the theory, both classically and after
quantisation is conformal, and will be referred to as conformal Toda
field theory or, simply, as Toda field theory. With the term $i=0$,
the conformal symmetry
is broken but the theory remains classically integrable, in the sense
that there are infinitely many independent conserved charges in
involution. The recent renewal of interest in Toda field theories
was stimulated by Zamolodchikov's ideas concerning perturbations of
conformal field theories
\NRF\rZd{\Zd\semi\HMa\semi\EYa\semi\FZa\semi\Ca\semi\Mb}
\NRF\rBCDSb\BCDSb\refs{\rZd ,\rBCDSb}.
The possible root systems which can be used in the lagrangian
\ltoda ,
maintaining the classical integrability are in one to one correspondence
with the untwisted and twisted affine Dynkin-Kac diagrams \refs{\rMOPa}.
However, in what
follows, it is useful to distinguish those which are unchanged (apart from
a possible flip) under the transformation
\eqn\duality{\alpha_i\rightarrow 2 \alpha_i /|\alpha_i|^2,}
and those which are \lq dual' pairs under this transformation. The self dual
set are $a_n^{(1)},\ d_n^{(1)},\ e_n^{(1)}$ whose roots are all of equal
length (conventionally, the longest root satisfies $|\alpha_i|^2$=2)
and $a_{2n}^{(2)}$; the dual pairs
are $(b_n^{(1)},a_{2n-1}^{(2)}),\
(c_{n}^{(1)},d_{n+1}^{(2)}),\  (g_2^{(1)},d_4^{(3)}),\
(f_4^{(1)},e_6^{(2)})$.

Each of the members of the self-dual set are untwisted
with roots of equal length, except for $a_{2n}^{(2)}$ which is twisted and
contains roots of three different lengths. The affine Toda theory corresponding
to the
simplest case $a_1^{(1)}$ is recognised to be the $\sinh-$Gordon theory (for
real coupling), or the sine-Gordon theory (for imaginary coupling, or
real coupling and imaginary fields).

Each of the non simply-laced or
twisted root systems can be obtained by \lq folding' one of the simply-laced
Dynkin or affine Kac-Dynkin diagrams, respectively. A  pair of
examples will suffice to illustrate this. Consider the Dynkin diagram for $d_4$
(first diagram).

\eqn\dfourgtwo{\dddiv1.4.3.2. \qquad\rightarrow\qquad \ddgd1.2.}

\noindent It has a symmetry
$\alpha_1\rightarrow\alpha_2\rightarrow\alpha_3\rightarrow\alpha_1$,
under which $\beta_1=\alpha_4$ and
$\beta_2 = (\alpha_1 +\alpha_2 +\alpha_3)/3$
are clearly invariant. The two roots $\beta_1, \beta_2$ are simple roots for
$g_2$
(that is, the second diagram, where the shorter root $\beta_2$
corresponds to the black spot). The extra root $\alpha_0=-(\alpha_1+\alpha_2
+\alpha_3+2\alpha_4)$ for $d_4$ is also invariant
and becomes the extra root $\beta_0=-(2\beta_1+3\beta_2)$ for $g_2^{(1)}$.
On the other hand, the dual
of $g_2^{(1)}$, $d_4^{(3)}$ is obtained using the threefold symmetry of
the extended Kac-Dynkin diagram of $e_6^{(1)}$ (first diagram).

$$\eddei1.6.2.3.4.5.0. \qquad\rightarrow\qquad \eddgiid1.2.0.$$

\noindent In this case, $\beta_1=\alpha_3$,
$\beta_2=(\alpha_2+\alpha_4+\alpha_6)/3$ and
$\beta_0=(\alpha_0+\alpha_1+\alpha_5)/3 =-\beta_1-2\beta_2$
are the invariant combinations under
the symmetry and provide the root system for $d_4^{(3)}$ (the second diagram,
in which the additional root attached to the $g_2$ Dynkin diagram is short).

The two types of $g_2$ extension lead to quite different classical field
theories.

\newsec{Classical integrability and classical data}

Toda field theory is classically integrable and indeed conformal
\NRF\rMa{\GNa\semi\Ma\semi\BCGTa}\refs{\rMa }. Affine
Toda field theory is classically integrable and also, in a generalised version,
conformal
\NRF\rBd{\BBa\semi\Bd}\refs{\rBd}.
Consider a conformal transformation in light-cone variables:
\eqn\lcconformal{x_\pm =(x^0\pm x^1)/2\rightarrow \bar x_\pm (x_\pm ).}
Clearly, since the second derivative of the scalar fields tranforms via
$$\partial_+\partial_-\phi\rightarrow
\bar\partial_+\bar\partial_-\phi ={\partial x_+\over\partial
\bar x_+}\, {\partial x_-\over\partial \bar x_-}
\ \partial_+\partial_-\phi ,$$
the equations of motion are invariant provided the potential term also
scales in a
suitable manner:
$$\sum_{i=1}^r n_i\alpha_i e^{\beta\alpha_i\cdot\phi}\rightarrow
{\partial x_+\over\partial \bar x_+}\, {\partial x_-\over\partial \bar x_-}
\ \sum_{i=1}^r n_i\alpha_i e^{\beta\alpha_i\cdot\phi}.$$
The latter requires the fields themselves to transform according to
\eqn\fconformal{\phi (x_+,x_-)\rightarrow\bar\phi (\bar x_+,\bar x_-)=
\phi (x_+,x_-) +{\rho\over \beta}\ \ln\left({\partial x_+\over\partial
\bar x_+}\, {\partial x_-\over\partial \bar x_-}\right),}
where the vector $\rho$ enjoys the property
\eqn\rhoproperty{\rho\cdot\alpha_i=1,\qquad i=1,2,3\dots ,r.}
Since the fundamental weights satisfy $$2\lambda_i\cdot\alpha_j/|\alpha_j|^2
=\delta_{ij},$$
$\rho$ may be expressed in terms of the fundamental weights:
$$\rho =\sum_{i=1}^r {2\over |\alpha_i|^2}\lambda_i.$$
It is immediately clear, since
$$\rho\cdot\alpha_0 =-\sum_{i=1}^r n_i=1-h,$$
that adding the extra term in the lagrangian
(proportional to $n_0$) breaks the conformal symmetry.

On the other hand, suppose the extra affine term is included, and further
suppose that the set of roots is actually taken to be the set of simple roots
for the affine algebra itself. Then, the set $\widehat\alpha_i,\ i=0,1,2\dots
,r$
are independent, lying in a Minkowski space of signature $(r+1,1)$
and, once again, a vector $\widehat\rho$ can be found for which
$$\widehat\rho\cdot
\widehat\alpha_i=1\qquad i=0,1,2,\dots ,r\ .$$
Using this, the argument of the last paragraph may be repeated to conclude
the theory is conformal even with the affine term included
\refs{\rBd}. The penalty
being paid for this is that the scalar fields $\phi$ no longer take values in
a Euclidean space and the energy is no longer a positive definite functional
of the field components. Restricting the fields to a Euclidean space
breaks the conformal invariance and, effectively, introduces a mass scale.

This situation is reminiscent of string theory which, in its most basic form,
contrives to describe
families of massive states starting from a conformally invariant lagrangian
whose fields take values in space-time.

Once conformal Toda field theory is quantised, it provides a coupling dependent
representation  of the Virasoro algebra whose central charge ($ade$ series) is
given by \refs{\rMa}:
\eqn\todac{c(\beta )=r+48\pi |\rho |^2 \left( {\beta \over 4\pi} +
{1\over \beta}\right)^2.}
This central charge is clearly symmetric under the transformation
$\beta \rightarrow 4\pi /\beta $, revealing that the quantum
conformal field theory enjoys a weak-strong coupling symmetry not
apparent in the original lagrangian.

Throughout these notes it will be assumed the fields take values in an
$r$-dimensional Euclidean space, spanned by the simple roots of the
Lie algebra $g$.

The classical integrability of the affine Toda field theories relies on
the existence of a Lax pair from which the conserved quantities may be
established. The details of this is a story in itself
\NRF\rOTd\OTd\refs{\rMOPa ,\rOTd} but from our
present perspective it is enough to be aware of some of the main results.
First of all, it is relatively straightforward to check the equivalence
between the zero
curvature property
$$F_{01}=\partial_0A_1-\partial_1A_0+[A_0,A_1]=0,$$
and the affine Toda field equations provided the two components of
the two-dimensional vector potential $A_\mu$ are given by:
\eqn\todalax{\eqalign{&A_0=H\cdot\partial_1\phi /2+\sum_0^r
m_i(\lambda E_{\alpha_i}-1/\lambda \ E_{-\alpha_i}) e^{\alpha_i\cdot\phi /2}\cr
&A_1=H\cdot\partial_0\phi /2+\sum_0^r
m_i(\lambda E_{\alpha_i}+1/\lambda \ E_{-\alpha_i}) e^{\alpha_i\cdot\phi
/2},\cr}}
where $H, E_{\alpha_i}$ and $E_{-\alpha_i}$ are the Cartan subalgebra and the
generators
corresponding to the simple roots and the extra root, respectively, of $g$.
Thus,
in particular,
$$\eqalign{[{\bf H},E_{\alpha_i}]&=\alpha_i\, E_{\alpha_i}\cr
[E_{\alpha_i}, E_{-\alpha_j}]&=\delta_{ij}\, {2\alpha_j\cdot{\bf H}
\over |\alpha_j|^2}.\cr}$$
The spectral parameter is $\lambda$, and the coefficients $m_i$ are chosen to
satisfy
$$m_i^2=n_i \alpha_i^2/8.$$
For convenience, the classically unimportant constants $m$ and $ \beta$ have
been
scaled away.

Since the path ordered integral of $A_1$,
$$T(a,b;\lambda )={\rm P}\, {\rm exp}\, \int_a^b dx^1A_1$$
satisfies
$${d\ \over dt}T=TA_0(b)-A_0(a)T\ ,$$
then
\eqn\tconservation{Q(\lambda )=tr\, T(-\infty ,\infty;\lambda )}
is time independent
when $\partial_1\phi \rightarrow 0$ as $x^1\rightarrow \pm \infty$
and $\phi (\infty )=\phi (-\infty ) +2\kappa$, where $\kappa\cdot\alpha_i$
is an integer.

An important fact about the Lax pair is the possibility of performing
a gauge transformation after which the potentials lie in a
Cartan subalgebra $h_i$
of $g$, two members of which are
$$E_{\pm 1}=\sum_{i=0}^r\, m_i E_{\pm\alpha_i}.$$
Once this gauge transformation has been done, the potential
$A_1$ takes the form
$$a_1=\lambda E_1 +\sum_{s\ge 1}\, \lambda ^{-s} h_s I_0^{(s)},$$
where the sum on the right hand side runs over the exponents of the
algebra $g$ (another characteristic set of integers which will be
met again below), modulo $h$, the Coxeter number. The elements of the
Cartan subalgebra are conveniently labelled by the $r$ exponents,
and $h_{s+nh}=h_s$. The zero curvature condition reads
$$\partial_0\, a_1 =\partial_1\, a_0\ ,$$
and therefore the integral of $a_1$ over the whole line is conserved.
Since $\lambda$ is arbitrary, there are infinitely many conserved quantities
$$Q_s=\int_{-\infty}^\infty dx^1\, I_0^{(s)}.$$
Adding or subtracting the equations \todalax\  reveals that $\lambda$
scales under a
Lorentz transformation ($\lambda\rightarrow l\lambda$)
in order to guarantee the correct transformation
of the light-cone
components of the vector potential. Consequently, the conserved quantities
$Q_s$ must scale under the transformation by a factor $l^s$.
(There is an alternative abelianisation for which there is a similar
expression for $a_1$ after the gauge transformation expressed as a
series of positive powers in $\lambda$. From this, a matching set of conserved
quantities of the opposite spin is obtained.)

It is possible to demonstrate the involutary nature of the charges
by first demonstrating the existence of a classical $r$-matrix for which
$$\{ T(\lambda )\  ,^\otimes\ T(\mu )\} \, =\, [r(\lambda /\mu ),
T(\lambda )\otimes T(\mu )], \qquad T(\lambda )\equiv T(-\infty , \infty
;\lambda ),$$
follows from the canonical equal-time Poisson bracket between the fields and
their
conjugate momenta. Indeed, Olive and Turok \refs{\rOTd} give $r$ in the form:
$$\eqalign{r(\lambda /\mu )={\mu^h+\lambda^h \over \mu^h-\lambda^h}
\sum_{i=1}^r H_i & \otimes H_i\cr
+{2\over \mu^h-\lambda^h}&\sum_{\alpha >0}\,{|\alpha|^2\over 2}
\left(\lambda^{l(\alpha )}\mu^{h-l(\alpha )}E_\alpha \otimes E_{-\alpha}
+\lambda^{h-l(\alpha )}\mu^{l(\alpha )}E_{-\alpha }\otimes E_{\alpha}\right)\
,\cr}$$
where the sum is over all positive roots of $g$ (ie all those roots expressible
as
combinations of simple roots with positive integer coefficients), and
$l(\alpha )$ is the length of a root (ie the sum of the integers in its
expansion
in terms of simple roots).

As mentioned above, the classically conserved charges are two dimensional
Lorentz tensors,
labelled by their \lq spin' in light-cone coordinates,  the possible
spins being the exponents of the algebra repeated modulo its Coxeter
number $h$. In other words, the conserved charges may be
denoted $Q_{s+kh}$, where $s$ is an exponent and $k$ is an integer. The
quantities $Q_{\pm 1}$ correspond to the light-cone components of the
energy-momentum. If the quantised field theory retains the
integrability property,
it is expected that the conserved quantities will survive
as mutually commuting
quantum operators whose eigenstates are the particles of the theory.
Thus, for single-particle states,
\eqn\qstates{Q_p|a>=q_p^a e^{p\theta_a}|a>\qquad p=s+kh,}
where $\theta_a$ is the rapidity of the particle labelled $a$:
\eqn\rapidity{p_a\equiv m_a(\cosh\theta_a,\sinh\theta_a),}
and $m_a$ is its mass.

Taking the classical lagrangian as the starting point for the definition
of a quantum field theory, the classical masses can be computed by
expanding the potential \vtoda\ as far as the quadratic term. Thus the
mass matrix is
\eqn\mmatrix{(M^2)^{ab}=m^2\sum_0^rn_i\alpha_i^a\alpha_i^b.}
For most cases, the mass matrix was diagonalised some time ago \refs{\rMOPa}.
However,
more recently, it was noticed
\NRF\rFKMa\FKMa \refs{\rBCDSb ,\rFKMa}
and then proved Lie algebraically by Freeman and others
\NRF\rFb{\Fb\semi\FLOa}\refs{\rFb}, that except
for the twisted cases
the eigenvalues of the mass matrix $m_a^2$ were themselves the squares
of the components of the lowest eigenvalue
eigenvector of the Cartan matrix corresponding
to $g$. In other words, it is possible to choose an ordering of the
masses so that $\v{m}=(m_1,m_2,\dots ,m_r)$ and
\eqn\masscartan{C\v{m}=4\sin^2{\pi\over 2h}\v{m}.}
This is quite a remarkable result since it allows the particles to be
assigned unambiguously (up to mass degeneracies),
to the Dynkin diagram for $g$. Curiously, the gravitational ordering
once this assignment is made follows the \lq weight' ordering in terms
of the dimension of the fundamental representations also
assigned to the spots on the Dynkin diagram. For example, the $a_n^{(1)}$
masses
($m_a=2m\sin{a\pi \over h}$)
increase from the ends of the Dynkin diagram working in, and are doubly
degenerate corresponding to the folding symmetry of the diagram;
for $e_8^{(1)}$
the masses are assigned as follows:

$$\ddeiii m_2.m_4.m_6.m_8.m_7.m_5.m_3.m_1.$$

Even more
remarkably, for the $ade$ series of simply-laced algebras (and for one
of the twisted cases $a^{(2)}_{\rm even}$), the classical mass ratios
are preserved in
perturbative field theory at least to one-loop order \NRF\rBCDSc\BCDSc
\NRF\rCMa{\CMa \semi\CMb}\refs{\rBCDSc ,\rCMa}, suggesting in turn
that the set of eigenvalues  $q_1^a$ in \qstates\ is an eigenvector of
the Cartan matrix for $g$. In a while, a generalisation
of this result will be discussed. The fact that the radiative corrections to
the
classical masses in most of the non simply-laced
cases are not universal is the first hint
that these theories will be rather different as quantum field theories.

Again at the classical level, it is interesting to examine the cubic
term in the expansion of \vtoda\ since this defines the classical
three-point couplings, needed to carry out for example the one-loop check
mentioned above. Once the mass eigenstates are known, it is
possible to compute the couplings,
$c^{abc}=\sum_0^rn_i\alpha_i^a\alpha_i^b\alpha_i^c$.
For many triples, the
coupling vanishes. However, when the coupling is not zero it is
proportional always\NRF\rBCDSc\BCDSc\NRF\rCMa{\CMa ,\CMb} \refs{\rBCDSc
,\rCMa}
to the area of a triangle whose sides have lengths
equal to the masses of the three participating particles $a,b,c$.
One consequence of this is that the coupling defines a set of angles
(the angles in the triangle),
by for example\foot{There is a convention in the
literature that the outside angles of the
triangle are denoted by $\theta_{ab}^c$, etc.},
\eqn\thetaabc{m_c^2=m_b^2+m_b^2-2m_am_b\cos\bar\theta_{ab}^c,}
where
\eqn\thetabar{\bar\theta =\pi -\theta .}
 Just which couplings are
non-zero will be explained further below once some of the geometry
of root systems has been explored.

It is tempting to suppose  eq\masscartan\ generalises (at least for the
simply-laced cases) and the other conserved
quantities have values constituting the components of the remaining
eigenvectors
of the Cartan matrix
\NRF\rKMa\KMa\refs{\rKMa}:

\eqn\cartaneigen{C\v{q}^{(s)}=4\sin^2{s\pi \over 2 h}\v{q}^{(s)}.}

This is true in the quantum theory, in the sense that it is consistent
with other known facts. Again, a fuller discussion is deferred.

\bigskip

\noindent{\bf Geometry associated with the Coxeter element}

\bigskip

There is some very useful geometry associated with roots and weights which is
less familiar than facts about representation theory. For that reason it will
be reviewed briefly here---further details may be found in several books,
for example, Bourbaki or Humphreys
\NRF\rBc{\Bc\semi\Hd}\refs{\rBc}.

A simple Weyl  reflection $w_i$ corresponds to a linear
transformation on the root lattice
representing a reflection in a plane orthogonal to the simple root $\alpha_i$
given by
\eqn\weylreflect{w_i: \qquad x\rightarrow x-2{\alpha_i\cdot x\over
\alpha_i^2}\alpha_i\ .}
A Coxeter element of the Weyl group is a product over the simple roots of the
simple
Weyl reflections. Clearly, once a set of simple roots have been chosen (ie a
set of
$r$ independent roots such that any other root is either a linear combination
of them
with positive coefficients, or a linear combination with negative
coefficients),
this product could be taken with different orderings of the individual simple
reflections.
However, different orderings lead to conjugate Coxeter elements. Alternative
choices of simple roots also lead to conjugate Coxeter elements. For  present
purposes, there is a special ordering which is extremely useful and which
relies on
the fact that Dynkin diagrams have no closed loops. The latter fact
allows the simple roots to
be divided into two sets such that the roots within each set are
mutually orthogonal (ie members of the same set are not joined by a line
in the Dynkin diagram). The members of the two sets are distinguished by
assigning
a colour to them, either black or white. Thus, for example, the $e_8$ diagram
can be coloured in this way as follows:
$$\ddeiii\W.\B.\B.\W.\B.\W.\B.\W.$$
The same is true of any other Dynkin diagram, as you can easily check\foot{It
is not true of all extended Dynkin diagrams, however; think of $a_{\rm
even}^{(1)}$}.
Obviously,
the product of Weyl reflections corresponding to simple roots within one of
these
special sets no longer matters since the Weyl reflections commute. With this
choice,
the Coxeter element is only ambiguous up to the relative black-white ordering,
and
for definiteness, the Coxeter element will be chosen for the rest of these
talks to be
\eqn\coxeterelement{w=w_\B w_\W \equiv \prod_{k\in\B}w_k\, \prod_{k\in\W}w_k .}
Notice, that each of the factors $w_\B$ and $w_\W$ separately squares to unity.
Notice, too, that there is a close relationship between the two factors of
the Coxeter element and the Cartan matrix of $g$:
\eqn\coxetercartan{(w_\B +w_\W )\alpha_i =\sum_j (2\delta_{ij}-C_{ij})\alpha_j\
.}
This is easily checked on the black and white roots separately. On the other
hand,
$$(w_\B +w_\W )^2 = 2 +w +w^{-1}\ ,$$
and  therefore
\eqn\ccsquare{(2 +w +w^{-1})\sum_ix_i\alpha_i =\sum_ix_i(2-C)^2_{ij}\alpha_j\
,}
revealing a close relationship between the eigenvectors of the Coxeter element
and the Cartan matrix. Indeed, the eigenvalues of the Cartan matrix have been
mentioned
already in connection with the classical data and, using them it is
straightforward
to deduce the eigenvalues of the Coxeter element. The eigenvalues of the
Cartan matrix are given in \cartaneigen , therefore the eigenvalues of the
Coxeter element are also labelled by the spins $s$ and are computed from
\ccsquare\ to be
$$e^{2i\pi s /h}.$$
Hence, the order of the Coxeter element is $h$.

To understand how the Coxeter element affects the roots, it is convenient to
define certain linear combinations of the simple roots whose coefficients are
the eigenvectors of the Cartan matrix.
Consider, for each spin $s$,
\eqn\asls{\eqalign{&a_\B^{(s)}=\sum_{i\in \B}q_i^{(s)}\alpha_i\qquad
l_\B^{(s)}=\sum_{i\in \B}q_i^{(s)}\lambda_i\cr
&a_\W^{(s)}=\sum_{i\in \W}q_i^{(s)}\alpha_i\qquad
l_\W^{(s)}=\sum_{i\in \W}q_i^{(s)}\lambda_i\cr}}
where the $\lambda_i$ are fundamental weights. Then,
\eqn\arelations{w_\B a_\B^{(s)}=-a_\B^{(s)}\ ,\qquad w_\B a_\W^{(s)}=
a_\W^{(s)}+2\cos\theta_s a_\B^{(s)}\ , \qquad \theta_s= {s\pi \over h}\ .}
The first of \arelations\ follows directly from the definition
of $w_\B$ and the
mutual orthogonality of the black roots; the second is less
straightforward and requires a
sequence of steps. Since the white roots have inner products with the black
roots
represented by the entries of the Cartan matrix,
$$\eqalign{w_\B a_\W^{(s)}&=a_\W^{(s)}-\sum_{{i\in\W}\atop
{j\in\B}}q_i^{(s)}C_{ij}
\alpha_j\cr
&=a_\W^{(s)}-\sum_{{i\in\W\cup\B}\atop {j\in\B}}q_i^{(s)}C_{ij}\alpha_j+
\sum_{{i\in\B}\atop {j\in\B}}q_i^{(s)}C_{ij}\alpha_j\cr
&=a_\W^{(s)}-\lambda^{(s)}a_\B^{(s)}+2a_\B^{(s)},\cr}$$
and the last line is the second relation in \arelations . Thus,
$$\eqalign{|w_\B a_\W^{(s)}|^2&=|w_\W a_\W^{(s)}|^2+4
\cos\theta_s a_\W^{(s)}\cdot
a_\B^{(s)}+4\cos^2\theta_s |a_\B^{(s)}|^2\cr
&=|a_\W^{(s)}|^2,\cr}$$
and there is a similar relation with black and white interchanged.
Comparing the two
leads to
\eqn\aproducts{|a_\W^{(s)}|^2=|a_\B^{(s)}|^2\qquad
\hbox{and}\qquad a_\W^{(s)}\cdot
a_\B^{(s)}=-\cos\theta_s |a_\W^{(s)}||a_\B^{(s)}|.}
Using the fact relating simple roots to fundamental weights, one also
has
$$\eqalign{a_\B^{(s)}&=\sum_{i\in\B\atop j\in\B\cup\W}
q_i^{(s)}C_{ij}\lambda_j\cr
&=2(l_\B^{(s)}-\cos\theta_s l_\W^{(s)}),\cr}$$
with a similar expression for $a_\W^{(s)}$. Hence,
\eqn\larelation{l_\W^{(s)}={a_\W^{(s)}+
\cos\theta_s a_\B^{(s)}\over
2\sin^2\theta_s}\ ,\qquad l_\B^{(s)}={a_\B^{(s)}+
\cos\theta_s a_\W^{(s)}\over
2\sin^2\theta_s},}
from which it is easily seen that
\eqn\lproducts{\eqalign{l_\B^{(s)}\cdot a_\W^{(s)}=&0=
l_\W^{(s)}\cdot a_\B^{(s)}\ ,
\qquad l_\B^{(s)}\cdot l_\W^{(s)}={\cos\theta_s\over 4\sin^2\theta_s}
|a_\B^{(s)}|^2\cr
&|l_\B^{(s)}|^2=|l_\W^{(s)}|^2={1\over 4\sin^2\theta_s}
|a_\B^{(s)}|^2.\cr}}
Clearly, all four vectors lie in a plane, for each choice of $s$.
The Coxeter element
acts as a clockwise rotation in this plane through an angle
$2\theta_s$. Notice,
that although it might happen that $a_\B^{(s)}$ and
$a_\W^{(s)}$ lie on the same
Coxeter orbit, this will never be the case for
$-a_\B^{(s)}$ and $a_\W^{(s)}$ (nor indeed for
$l_\B^{(s)}$ and $l_\W^{(s)}$).
The various vectors are illustrated in the
diagram below.

\epsffile{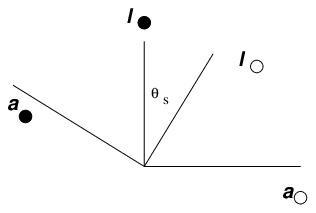}
\vskip -12pc
\centerline{Vectors in the $s$ plane}

The sets $w^p\alpha_i, \ i\in\W$ and $-w^p\alpha_i, \ i\in\B$, for
$p=1,\dots ,h$, of images of simple roots do not intersect for distinct
simple roots. They provide rank $r$ orbits each of $h$ elements, together
providing the full set of roots for the algebra $g$.
Since it is possible to normalise the eigenvectors of the Cartan matrix
so that
$$\sum_s q_i^{(s)} q_j^{(s)}=\delta_{ij},$$
the relationship between $a_\W^{(s)}$ and $a_\B^{(s)}$ can be
inverted to find that
each simple root has a component in the spin $s$ plane,
either along $a_\B^{(s)}$
or along $a_\W^{(s)}$, according to its colour. Moreover, the
images of each simple root
$\alpha_i$ under repeated application of the Coxeter element
each have a component
in this plane of the same magnitude, equal to $q_i^{(s)}$.
In particular, for
$s=1$, and according to the earlier observation \masscartan ,
each orbit has a classical mass associated with it and
therefore the whole orbit
may be assigned to a particular particle.

Consider three roots which make a triangle. The projection
of this triangle onto the $s=1$
Coxeter plane provides another triangle the sides of which
 have lengths
equal to the masses of the particles
associated with the three orbits to which the three
roots belong.

More interestingly,
it is now possible to give a characterisation of the
three-point couplings:
\bigskip
{\bf The coupling $c^{abc}$ between three
particles $a,b$ and $c$ is non-zero
if and only if
there are three vectors, one from each of the orbits
representing the particles, which sum to zero}.
\bigskip

\noindent This was first
proved on a case by case basis by Dorey
\NRF\rDc{\Dc\semi\Dd}\refs{\rDc}
and then
deduced from the classical lagrangian
by Fring, Liao and Olive
by extending the ideas of Freeman \refs{\rFb}. For all cases,
the couplings actually correspond to a subset of the Clebsch-Gordan series in
the
sense that if a coupling $abc$ is non-zero then the tensor product of the
representations
assigned to the particles according to their assignment to the Dynkin diagram
contains the
trivial representation, ie ${\bf a\otimes b\otimes c \supset 1}$. Except for
the cases corresponding to
$a_n^{(1)},\ d_4^{(1)}$ the converse is not true. The relationship between the
couplings
and the Clebsch-Gordan series, and other matters, has been further elucidated
by Braden
\NRF\rBa\Ba\refs{\rBa}.

There is another way to label the Coxeter orbits
\NRF\rKb\Kb\refs{\rKb}
which will turn out to be useful in
the next section, and which naturally incorporates
the minus sign. Let the elementary Weyl
reflections and the roots be labelled so that
$$w=w_\B w_\W =w_1\dots w_b w_{b+1}\dots w_r,$$
then set
\eqn\phis{\eqalign{\phi_k&=w_r w_{r-1}\dots w_{k+1}\alpha_k\cr
&=\cases{\alpha_k\ & for\ $k\in\W$\cr
w_\W\alpha_k =-w^{-1}\alpha_k &for\ $k\in\B$\cr}\cr
&=(1-w^{-1})\lambda_k.\cr}}
The last two lines of \phis\ follow directly from the definition
in the first. One consequence
of the second fact is that the set of images of distinct $\phi_k$
never overlap and,
therefore, these vectors may be used equally well to label the
orbit which has been
associated with a particle. A curious property of these
vectors, which turns
out to have a use in the next section, is the following.
The image of each of them
under the inverse Coxeter element is a positive root and
successive images remain positive
until the middle of the orbit, after which they all change sign,
 remaining
negative subsequently for the rest of the orbit.

 As an illustration, consider $d_4$
labelled as before, with the outer spots coloured black and the
centre spot white. Then
$$\phi_k=\alpha_k+\alpha_4\ \hbox{for}\  k=1,2,3\qquad
\phi_4=\alpha_4,$$
and the orbits of the inverse Coxeter element are
\eqn\dfourorbits{\eqalign{1:\quad&\alpha_1+\alpha_4;\
\alpha_2+\alpha_3+\alpha_4;\ \alpha_1;
\ -\alpha_4-\alpha_1;\dots \cr
2:\quad&\alpha_2+\alpha_4;\ \alpha_1+\alpha_3+\alpha_4;\
\alpha_2;\ -\alpha_4-\alpha_2;\dots \cr
3:\quad&\alpha_3+\alpha_4;\ \alpha_2+\alpha_1+\alpha_4;\
\alpha_3;\ -\alpha_4-\alpha_3;\dots \cr
4:\quad&\alpha_4;\ \alpha_1+\alpha_2+\alpha_3+2\alpha_4;\
\alpha_1+\alpha_2+\alpha_3+\alpha_4;
\ -\alpha_4;\dots .\cr}}
Clearly, these orbits provide the full set of roots as promised.

\newsec{Aspects of the quantum field theory}

In this lecture, the intention is to provide certain basic
facts and formulae
which have proved to be remarkably universal. Most of the
time, the $ade$ sequence of theories will be considered.
The affine diagrams for these
are invariant under the transformation \duality . For background,
and further references on S-matrix theory in two dimensions,
the review article by Zamolodchikov and Zamolodchikov
\NRF\rZZa\ZZa\refs{\rZZa}
is strongly recommended.

It will
be supposed, as a working
hypothesis, that (1) after quantisation the conserved charges
remain conserved and in
involution---ie commute with one another, and (2) the particle
spectrum of any one
of these theories is as
simple as possible---in other words, the particles are exactly
$r$ in number, stable and
distinguishable, if
not by their masses then by one or other of the  conserved
charges\foot{The sine-Gordon
model is not as simple as this. There are two \lq soliton'
states which are only distinguished
by a zero spin charge. Such a distinction is already too
relaxed for the present purposes;
it permits the mixing of soliton and anti-soliton in a
scattering process, leading, in turn,
to a greatly enriched spectrum of bound states.}.
With this in mind, it will be assumed that there
is a set of one particle states which are eigenstates of
the quantum conserved charge operators
(which have not been properly constructed yet), ie \qstates :
\eqn\oneparticle{Q_p|a>=q^p_a e^{p\theta_a}|a>\qquad p=s+kh,}
where $\theta_a$ is the rapidity of particle $a$.
In addition two-particle states are also assumed to be
 eigenstates of the conserved charge
operators, ie:
\eqn\twoparticle{Q_p|a,b>=(q^p_a e^{p\theta_a}+q^p_b
e^{p\theta_b})|a,b>,}
and so on.

There is no elementary definition of these particle states,
although they may be approximated
perturbatively. If it is further supposed that two-particle
states, which are functions of a pair
of rapidities, one for each particle, may under certain
circumstances be dominated by a
single particle state, then
\eqn\conservation{q^p_a e^{p\theta_a}+q^p_b e^{p\theta_b}=
q^p_{ \bar c}e^{p\theta_{\bar c}},}
where the particle $ \bar c$ must itself be part of the
conjectured spectrum. If the
particle $c$ is to be stable then this situation  cannot
occur for real rapidity
difference $\Theta_{ab}=\theta_a-\theta_b$. Rather,
considering the spin $\pm 1$ charges
(the light-cone components of energy-momentum,
$q^{\pm1}_k=m_k$), the situation may arise
only when the rapidity difference satisfies
\eqn\schannel{m_{\bar c}^2=m_a^2+m_b^2+m_am_b
\cosh\Theta_{ab}=m_a^2+m_b^2+
m_am_b\cos U_{ab}^{ c},}
and the masses $m_a,m_b$ and $m_c$ are the sides of a
triangle with internal angles $\bar U_{ab}^c,
\bar U_{ac}^b,\bar U_{bc}^a$. The same triangle equally
well permits a
description of the energy-momentum
conservation for the virtual processes
$ac\rightarrow \bar b$ and $bc\rightarrow \bar a$.
For these special rapidity differences,
the rapidities themselves may be written
conveniently as
\eqn\rapidities{\theta_a=\theta_{\bar c}-i
\bar U_{ac}^b\qquad \theta_b=\theta_{\bar c}+i\bar U_{bc}^a.}

One might expect that for a certain rapidity difference
the vacuum state may dominate a particle-
anti-particle state. For this, energy momentum requires
$\Theta_{a\bar a}=i\pi$ and therefore,
\eqn\antiparticle{q^p_ae^{p\theta_a}+
q_{\bar a}^pe^{p(\theta_a-i\pi )}=0 \qquad\hbox{ie}\qquad
q_{\bar a}^p=(-)^{p+1}q^p_a.}
One consequence of this is immediate. Particles and
anti-particles are distinguished only by
even spin charges. Affine Toda theories for which the
exponents are odd must contain
self-conjugate particles only (this includes $e_7^{(1)}$ and
$e_8^{(1)}$ which have
no mass-degenerate states,
and $d_{\rm even}^{(1)}$ which has mass degenerate states
corresponding to the prongs of
the fork in the Dynkin diagram).

More generally, using \antiparticle , the
$ab\rightarrow \bar c$ conserved charge
relation may be rewritten
\eqn\qtriangle{q^p_ae^{ip(U_{ac}^b+U_{bc}^a)}+
q^p_ce^{ipU_{bc}^a}+q^p_b=0,}
which represents a series of \lq triangular '
relations, one for each $p$.

At this stage, it is tempting to identify the
set of triangle relations \qtriangle\ with
the projections of the root triangles which
represent the classical couplings described
in the last section \refs{\rDc}. This would require the
masses of the particles in the quantum spectrum
to be essentially the same (upto an overall
scale) as the mass parameters in the classical
lagrangian; the coupling angles $U_{ab}^c$ would
also be the same as those for the
classical mass triangles \thetaabc . The eigenvalues
of the conserved quantities $q^p_a$
would repeat modulo $h$, and the first $r$ of them,
labelled by the exponents of the
algebra, would be the components of the corresponding
Cartan eigenvector \cartaneigen .
These consequences of the initial hypotheses are very
strong and would need to be verified
by  direct calculation. In fact, if one examines the
members of the self-dual affine
Toda theories perturbatively, all infinities may be
removed by normal-ordering and
a  calculation of the \lq bubble' diagrams which
contribute to mass corrections at lowest order
reveals that the identification of the classical
masses with the quantum masses is natural
in the sense that the mass corrections are independent
of particle type. This is
quite definitely not the case for those theories which
have a different dual partner.
For them, the mass corrections are type-dependent and
it would seem unnatural to
insist on the quantum masses being the same as the
classical ones. In the next
lecture an alternative and more attractive approach
to these will be presented.

Given the large number of conserved charges and the
set of distinguishable
particles, the two particle scattering of affine
Toda particles is expected to
be simple in the sense that the character of the
particles is unchanged, there is no
production, and the initial and final momenta are
the same \refs{\rZZa}.
Indeed, the \lq in' and \lq out' states may differ
only by a phase
which may at most (because of Lorentz invariance)
be a function of the rapidity
difference of the two particles and of the coupling
$\beta^2$ (or $\hbar$). Ie
\eqn\smatrix{|a,b>_{\rm out}=S_{ab}(\Theta_{ab};\beta )
|a,b>_{\rm in}.}
For each pair of particles there will be such a phase
factor. The set of
factors will be called the two-particle S-matrix
although there is no real
scattering going on. The phase factors regarded as
functions of complex
rapidity difference are far from trivial, however.
Indeed, they are analytic functions
of the rapidity difference\foot{Using the rapidity variable
 effectively removes the
$s,t$ threshold cuts and there are no others because
there is no production.}, with an
intricate set of zeroes and poles characteristic of
each specific theory. The
\lq physical strip' consists of the region
$0<Im(\Theta )<i\pi$, the boundary
$Im(\Theta )=0$ being  the physical $s-$channel,
 the boundary $Im(\Theta )=i\pi$
being the physical $t$-channel. The region
$0>Im(\Theta )>-i\pi$ is the second,
unphysical sheet from the point of view of the
Mandelstam variables.  The
continuation of the unitarity and crossing
relations away from $Im(\Theta )=0$
requires
\eqn\unitarity{S_{ab}^{-1}(\Theta )=S_{ab}(- \Theta )
\qquad \hbox{and}\qquad
S_{a\bar b}(\Theta )=
S_{a b}(i\pi -\Theta ),}
respectively. Taken together, these imply the
S-matrix elements are $2\pi i$ periodic.

If it is further assumed that the scattering is
factorisable (one more
feature which will need substantiating ultimately),
then the three-particle
S-matrix elements may be regarded as products of
two-particle S-matrix elements.
The ordering ambiguity (normally resolved by the Yang-Baxter
equation) is absent here
because of the special nature of the two-particle
scattering (there is no reflection).
Thus, the Yang-Baxter equation itself plays no r\^ole.
On the other hand, a two-particle
state for complex rapidity difference  may share
the quantum numbers of a
single particle state. The signal for this is a
direct channel pole
in the physical strip
at a purely imaginary rapidity difference. The
fusing idea
allows a set of \lq bootstrap' relations to be
formulated which relates the scattering
of particle $d$, say, with $a$ and $b$, to the
scattering of $d$ with $\bar c$. Ie,
pictorially,

\epsffile{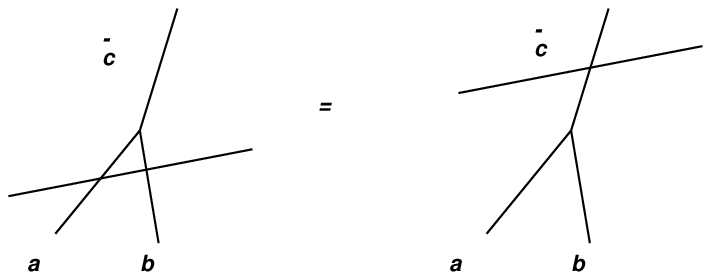}

\noindent and algebraically
\NRF\rKa\Ka\refs{\rKa}:
\eqn\bootstrap{S_{d\bar c}(\Theta )=S_{da}
(\Theta -i\bar U_{ac}^b)S_{db}
(\Theta +i\bar U_{bc}^a).}
The latter, in the case of the two particle
state $a,\bar a$ at a
relative rapidity of $i\pi$, is in agreement with the crossing
relation.

Using \unitarity , \bootstrap\ can be rearranged to
\eqn\bootstrapprod{S_{da}(\Theta +i U_{ac}^b+iU_{bc}^a)
S_{dc}(\Theta +iU_{bc}^a)S_{db}(\Theta )=1,}
which is a \lq product' version of the sum rule \qtriangle .

The equation \bootstrap\ is extremely useful but it does not
fix the S-matrix elements uniquely. What it does do is supply
a set of consistency conditions which must be supplemented by
other data or prejudices. A natural idea,
given the hypothesis concerning
the masses of the particles, is to suppose
that the possible fusings
for which the bootstrap works are to be given
precisely by the classical couplings
$c^{abc}$ and their associated angles \refs{\rAFZa ,\rBCDSb ,
\rBCDSc ,\rCMa}.
Before checking this, however,
it is also necessary to make some remarks
concerning the coupling dependence
of the S-matrix elements.

Clearly, when $\beta =0$ the S-matrix elements
ought to be unity since the particles
are free. When $\beta\ne 0$, the poles indicating
the fusings are at fixed positions
and these must  be the only poles on the
physical strip since it has
been presumed that the classical spectrum is complete.
Therefore, the S-matrix
elements must contain travelling zeroes on the physical
 strip which coincide with the
fixed poles to cancel them when $\beta =0$. Because
of unitarity, each zero
has an accompanying pole which must be situated on
the unphysical strip for
any choice of the coupling in the range $0\le\beta\le\infty$.
For the simply-laced
conformal Toda theories, it was pointed out (eq\todac ) that there
is a symmetry between strong
and weak coupling in the sense that the Virasoro central
charge is
actually invariant under  $\beta \rightarrow 4\pi /\beta$.
There is an elegant solution to
the bootstrap which also enjoys this symmetry and which
neatly parametrises the
coupling dependence to ensure the other desirable properties.

It is useful to have a convenient notation for the basic
ratio of functions
satisfying the periodicity and unitarity relations
\refs{\rBCDSc}. Set
\eqn\basicblock{(x)_\Theta =\sinh\left({\Theta\over 2} +
{i\pi x\over 2h}\right)\biggm/
\sinh\left({\Theta\over 2} -{i\pi x\over 2h}\right),}
bearing in mind that the fusing angles are always
multiples of $\pi /h$.
Often, this will be referred to merely as $(x)$.
The fixed poles will be represented by
terms of this kind. The coupling dependence may be
incorporated by assembling blocks
of this type as follows:
\eqn\bigblock{\{ x\}_\Theta = {(x-1)(x+1)\over (x-1+B)(x+1-B)},}
where $B$ is coupling dependent and, in fact, universal,
\eqn\Bdef{B(\beta )={1\over 2\pi}{\beta^2\over 1+4\pi /\beta^2}.}
Clearly, for small $\beta$, $\{ x\}_\Theta$ approaches
unity and, because
$B(\beta )=2-B(4\pi /\beta )$, for large $\beta $,
exactly the same is true; the
pole-cancelling zeroes
cross over and cancel the opposite pole, as $\beta$
runs from $0$ to $\infty$. In principle,
other functions of $\beta$ might be acceptable under
these constraints but this is the
one originally suggested by Arinshtein, Fateev and
Zamolodchikov \refs{\rAFZa} for the $a_n^{(1)}$
series, on the basis of a comparison with the
sin/sinh-Gordon model \refs{\rZZa}.

Rather than simply writing down the conjectures
for the S-matrices, it is instructive to build
one up, watching the bootstrap in action. An
interesting case is $d_4^{(1)}$ labelled as in
fig\dfourgtwo . There are four distinguished
particles 1,2,3 with mass $\sqrt{2}m$ and 4 with mass
$\sqrt{6}m$, and possible couplings
$$c^{123}\qquad c^{aa4}\ (a=1,2,3) \qquad
\hbox{and }\qquad c^{444}.$$
In this case, the Coxeter number $h=6$.
The particles are self-conjugate and therefore
the S-matrix elements are crossing symmetric.

To begin with, make the simplest compatible
conjecture for $S_{12}$, say. It ought to have a
pole at $\Theta =2i\pi /3$ with a positive residue,
and a crossed partner at $\Theta =i\pi /3$
with negative residue, and no other fixed poles.
In the above notation, a reasonable guess would
be
\eqn\dfoursi{S_{12}( \Theta )=\{ 3\}\equiv
{(2)(4)\over (2+B)(4-B)}\sim {i\over \Theta -
2\pi i/3}\ {\pi B\over 6}\qquad \left(\ =
S_{13} =S_{23}\ \right)\ .}
Using this pole together with the bootstrap
\bootstrap , leads to
\eqn\dfourii{\eqalign{S_{33}(\Theta )& =
S_{13}(\Theta -i\pi /3)\, S_{13}(\Theta +i\pi /3)
\cr & =\{1 \}\{ 5\}
\equiv {(0)(2)(4)(6)
\over (B) (2-B)(4+B)(6-B)}\qquad
\left( =S_{11} =S_{22}\ \right)\ ,\cr}}
which is again crossing symmetric.
However, because $(6)=-1$, it is the pole at
$\Theta =i\pi /3$ which has the positive
residue this time, indicating the
non-zero coupling $c^{334}$ (or $c^{114},
\ c^{224}$). Using the latter with the bootstrap
yields
\eqn\dfouriii{\eqalign{S_{14}(\Theta )&=
S_{13}(\Theta -i\pi /6)\, S_{13}(\Theta +i\pi /6)\cr
&=\{ 2\}\{ 4\}\equiv {(1)(3)^2(5)\over (1+B)(3-B)(3+B)(5-B)}\qquad
\left( =S_{24} =S_{34}\ \right)\ ,\cr}}
and
\eqn\dfouriv{\eqalign{S_{44}(\Theta )&=
S_{14}(\Theta -i\pi /6)\, S_{14}(\Theta +i\pi /6)\cr
&=\{ 1\}\{ 3\}^2\{ 5\}\equiv -{(2)^3(4)^3\over
(B)(2-B)(2+B)^2(4-B)^2(4+B)(6-B)}\ .
\cr}}
All other bootstrap relations are verified by direct checking.

The S-matrix elements may be represented conveniently in the following
diagram.

\epsffile{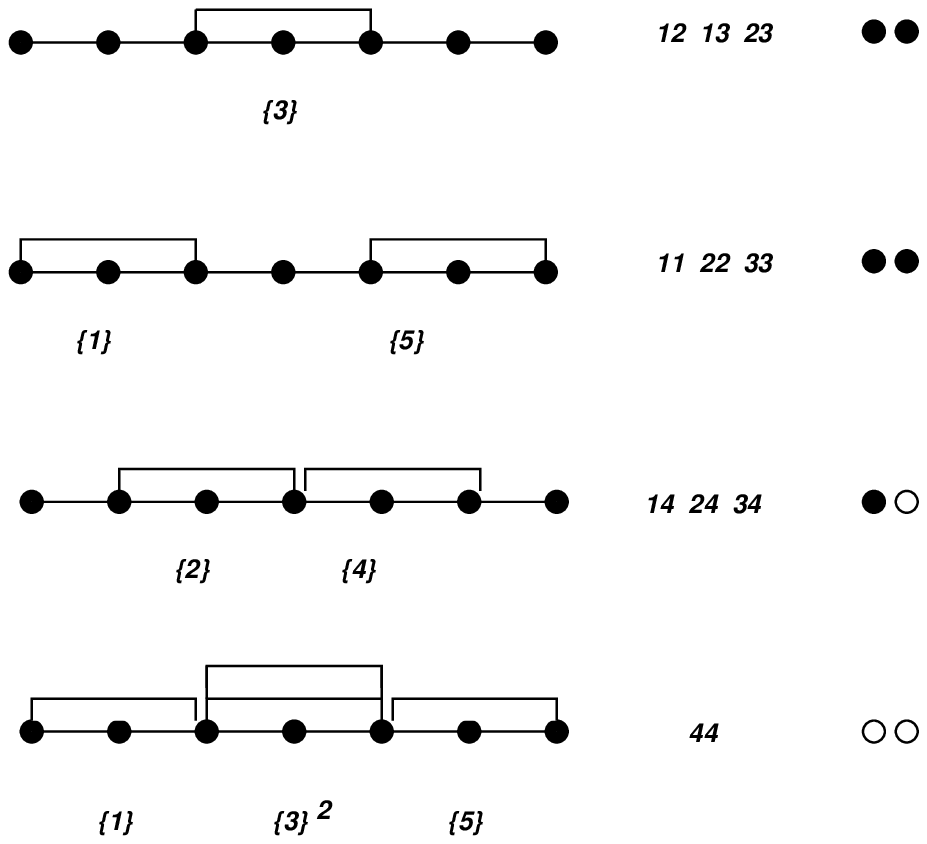}
\vskip -3pc
\noindent In the diagram, the physical strip is marked at intervals of $\pi /6$
and the boxes represent the basic block factors in the S-matrix
elements whose labels appear on the right. The vertical lines represent the
fixed pole positions on the physical strip.

Now, return to the Coxeter element orbits provided in \dfourorbits .
There is a striking correlation between the coefficient of $\alpha_i$
in the expressions for the vectors in the positive
part of the orbit of $\phi_j$ ( ie
the vectors $w^{-p}\phi_j$ for $p=0,1,2$),
and the boxes which appear in the diagram representing $S_{ij}$.
Indeed, the boxes are labelled by $2p+1+\epsilon_{ij}$, where $\epsilon_{ij}$
depends only on the colour of the pair $i,j$:
\eqn\epsilonij{\epsilon_{\B\B}=\epsilon_{\W\W} \qquad \epsilon_{\W\B}=
-\epsilon_{\B\W}=1.}
This observation suggests there is a formula for the S-matrix elements
in terms of roots weights and Coxeter orbits
\refs{\rDc}:
\eqn\doreys{S_{ab}(\Theta )=\prod_{p=1}^h
\{ 2p+1+\epsilon_{ab}\}^{\lambda_a\cdot w^{-p}\phi_b}_+ .}
(The $+$ subscript indicates that because the blocks are all
accounted for by traversing the positive part of the orbit of
$\phi_b$ only, it is necessary, when extending the product over the
whole Coxeter orbit, to realise that the numerators of the blocks
are reconstructed by the positive part of the orbit and the denominators
by the negative part.)
Because the formula depends only on the roots/weights it promises to
be universal. Actually, that is indeed the case. The S-matrix formula
once postulated can be shown to be symmetrical under
$a\leftrightarrow b$, to be unitary, to satisfy crossing and to
satisfy the bootstrap relation
\NRF\rFOa\FOa\NRF\rDf\Df\refs{\rDc ,\rFOa ,\rDf}.
This is a beautiful result, which
applies only to the $ade$ series of cases, and it is a pity
there is no direct derivation of it from the field theory.

The observant will have noticed that among the poles in the S-matrix
elements \dfouriii\  and \dfouriv\ there are some of order two and three.
These are clearly required by the bootstrap and are, in a sense,
fortuitously useful. The point is that there is little hope of
computing directly the S-matrix elements perturbatively, at least
for arbitrary rapidity, but there is some hope of calculating the
coefficients of higher order poles. This is because the poles appear as
Landau singularities in Feynman diagrams and there is a well
developed calculus for dealing with them. For example, there is no
time to go into details, but the double poles all arise from
singularities of box diagrams and it has been checked that the
coefficients of the poles to order $\beta^4$ agree with
the predictions of the S-matrix elements (not just for $d_4^{(1)}$,
but in all cases). This is quite important because the observant
will also have noticed that there was no attempt to check
the bootstrap on the double pole in \dfouriii . The fact that the
poles are an artefact of the perturbation expansion and there is
no order $\beta^2$ tree-graph with a simple pole sharing the same pole
position strongly
suggests this is a correct interpretation of the bootstrap rules.
On the other hand, third order poles (and in general odd-order poles)
appear as dressings of tree processes and one would expect that
their existence does really signal a bound state which ought
to participate in the bootstrap. It has been found that there are
a number of diagrams of different types contributing to the third
order poles (all two-loop diagrams, since the leading contribution to
the third order pole is order $\beta^6$), but never more than
twenty-six (!) as one ranges over the $ade$ series. The sum of the
contributions from these diagrams agrees exactly with the prediction
from the conjectured S-matrix elements whenever the cubic
poles occur in the
$ade$ series. The number of diagrams to be computed is prohibitively
large for the fourth and higher poles (up to order twelve in the
$e_8^{(1)}$ S-matrix elements), and for these a direct check is not
possible. The type of checking advocated here is complicated and
makes it abundantly clear how inefficient the perturbation series
is from a computational point of view. These poles have been checked to order
three
in
\NRF\rBCDSe\BCDSe\refs{\rBCDSe}, and other perturbative matters have been
investigated in \NRF\rBSa{\BSa\semi\BSb\semi\BCKKSa\semi\SZb}\refs{\rBSa}.

\newsec{Dual pairs}

The theories based on non simply-laced algebras work in a very different way
which will be partially explained by reference to a particular example.
Further details are obtainable in the recent literature
\NRF\rDGZa\DGZa
\NRF\rCKKa\CKKa
\NRF\rKWa\KWa
\NRF\rWWa\WWa
\NRF\rCDSa\CDSa
\NRF\rDe\De\refs{\rDGZa -\rDe} although there
is much yet to do before the final version of the story can be told.

For definiteness, consider the pair of classical theories based on
the  extended Dynkin diagrams $g_2^{(1)}$ and $d_4^{(3)}$:
$$\eddgid0.1.2.$$
$$\eddgiid1.2.0.$$
In each case, the black spots denote short roots and it is clear that
to obtain one diagram from the other involves an inversion of roots \duality .

In each case, there are two particles labelled 1 and 2 but their
mass ratios are different \refs{\rBCDSc ,\rCMa}. In the first case, the
classical mass
parameters are simply those of $d_4^{(1)}$ without the mass degeneracy,
since this has been removed by the folding corresponding to the threefold
symmetry of the $d_4$ Dynkin diagram. In the second case, the root system
is obtained by applying the folding procedure to the extended Dynkin diagram
$e_6^{(1)}$, which also has a threefold symmetry; hence, in this case the
masses are a subset of those to be found in the $e_6^{(1)}$ theory.
In summary,
the two mass ratios are
\eqn\gtwomass{{m_1\over m_2}\biggm|_{g_2^{(1)}}={\sin (\pi /6)\over
\sin (2\pi /6)}\qquad {m_1\over m_2}\biggm|_{d_4^{(3)}}={\sin (\pi /12)\over
\sin (2\pi /12)}.}
Moreover, the non-zero three-point couplings for the two cases are:
\eqn\gtwocoupling{g_2^{(1)}:\  111,\ 112,\ 222,\qquad d_4^{(3)}:\  111,
\ 112,\ 222, \ 221\ .}
Hence, from a classical point of view these two theories are very different.

Some time ago, it was also noted that guesses for the S-matrix for
$g_2$ were problematical if based on maintaining poles at the positions
of the classical masses \refs{\rBCDSc}. There were always extra singularities
whose
origin could not be traced in perturbation theory. It was also found that
radiative mass corrections, which worked very well for the simply-laced
theories, did not preserve the classical mass ratios, suggesting
either that  cases such as $g_2$ were in a sense anomalous and therefore
not quantum integrable or, that they were quantum integrable but that the
relationship with the classical theory was much less clear cut.
The principle step in suggesting a resolution of these difficulties
has been provided by Delius, Grisaru and Zanon \refs{\rDGZa}. They have noted
how
the bootstrap might be satisfied, even in a situation where there are
particles with coupling dependent masses, in such a manner that the
small coupling approximation is provided by the $g_2^{(1)}$ theory
and the large coupling limit is provided by the ${d_4^{(3)}}$ theory.
In other words, there is a quantum field theory corresponding to the
pair together rather than either classical theory separately, and the
transformation
$$\beta\rightarrow 4\pi /\beta ,$$
effectively implements the inversion \duality\  which interchanges the
two extended Dynkin diagrams. A similar mechanism is working for all the
non simply-laced algebras which come in the pairs listed previously
and related by \duality . The exceptions to this are the members of the
$a_{2n}^{(2)}$ sequence which are \lq self-dual'.

The first thing to note is that the masses may be parametrised conveniently
by setting
\eqn\betamass{{m_1\over m_2}\biggm|_{\beta}={\sin (\pi /H(\beta ))\over
\sin (2\pi /H(\beta ))}\qquad \hbox{with}\qquad 6\le H(\beta )\le 12
\qquad \hbox{for}\qquad 0\le\beta \le\infty ,}
where the functional dependence of $H$ on $\beta$ is really a matter of
informed conjecture. A few words will be said about it at the end of the
section.
That the masses do depend on the coupling has been confirmed by
Watts and Weston \refs{\rWWa} who have investigated the coupling dependence
in a Monte-Carlo lattice simulation of the model. Their results leave
little doubt that the masses do indeed flow with the coupling although
the numerical accuracy of the simulation is not yet sufficient to pin down
the actual dependence on $\beta $.

The couplings \gtwocoupling\ are more problematical since
the two theories have different numbers of three-point couplings. However,
the two self-couplings are clearly permitted whatever the masses might be
and always correspond to a coupling angle of $2\pi /3$ in the notation
introduced before (eq\thetaabc ). Also, the $112$ coupling is quite
natural with a
coupling angle of $2i\pi /H$, since
$$\sin^2\left( {2\pi\over H}\right)\equiv 2\sin^2\left({\pi\over H}\right)
+2\sin^2\left( {\pi\over H}\right)\cos\left( {2\pi\over H}\right),$$
whatever the value of $H$ might be, whereas the coupling 221 is
quite unnatural. As far as an $ab$ S-matrix element
is concerned, one would expect the $abc$
couplings to emerge as poles (or possibly multiple poles) in the
physical strip  with a positive coefficient (times $i$). That was
certainly what happened in the simply-laced sequences of models.
However, in this and other similar cases, the mere positivity of the pole
coefficient is not enough and it appears to be necessary to strengthen
the requirement to {\bf positivity over the whole range of $\beta$}.
Once this is done it is found that there is a consistent set of
bootstrap conditions satisfied by a subset of the classical couplings,
but not all of them.

To examine the S-matrix, it is helpful to use a diagrammatic
representation (see below) which displays the poles on the physical strip
(solid lines) and compensating zeroes (dashed lines), as they travel from
their positions at $\beta =0$ to their positions at $\beta =\infty$. The
filled circles represent points on the physical strip at intervals of $
\pi /h$ or $\pi /h^{\vee}$. Thus, the upper row represents the physical strip
marked at intervals of $\pi /6$ for $g_2^{(1)}$ while the lower row
represents the physical strip marked at intervals of $\pi/12$ for $d_4^{(3)}$.
The dashed lines always meet solid lines at the top and bottom of the diagram
indicating that the poles and zeroes precisely cancel there, as they ought
because the S-matrix elements are unity at $\beta =0$ or $\infty$.
The first of the diagrams represents $S_{11}(\Theta  )$

\epsffile{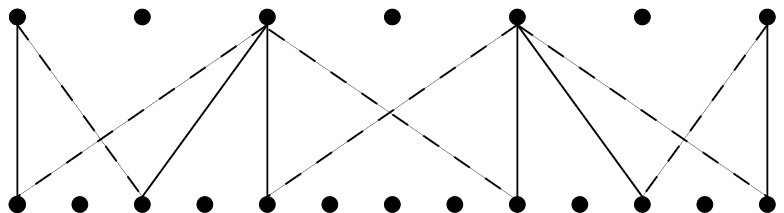}
\centerline{$S_{11}(\Theta  )$}
\bigskip
\noindent for which the algebraic expression is
\eqn\gtwosone{ {(0)\ (2)\over (H/3 -2)(4- H/3)}\, {(H/3)\ (2H/3)\over
(4) (H-4)}\, {(H-2)\ (H)\over (2+2H/3)(4H/3 -4)}\ ,}
where the bracket notation has been adjusted to represent
$$(x)=\sinh\left({\Theta\over 2}+{xi\pi\over 2H}\right)\biggm/
\sinh\left({\Theta\over 2}-{xi\pi\over 2H}\right).$$
It is clear from the diagram that there are two physical simple poles
and their crossed partners---the third vertical line represents
the self coupling $11\rightarrow 1$,
and the oblique line next to it represents the $11\rightarrow 2$ coupling.
Using the bootstrap relation on the $11\rightarrow 2$ coupling leads directly
to the S-matrix element $S_{12}(\Theta  )$ for which the algebraic expression
is
\eqn\gtwostwo{ {(1)\ (2H/3 -1)\over (H-5)(5-H/3)}\, {(H/3+1)(H-1)\over
(4H/3-5)(5)}.} and which is represented diagrammatically by

\epsffile{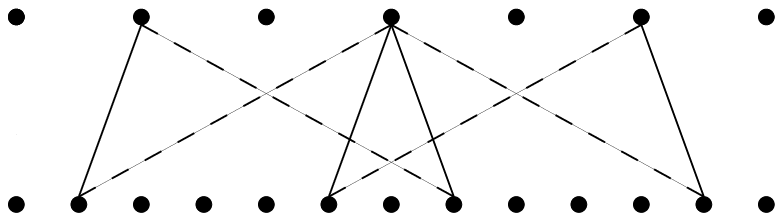}
\centerline{$S_{12}(\Theta  )$}

\noindent
In this case, the rightmost solid line represents the physical pole for
the expected bound state in the channel $12\rightarrow 2$ and it looks as
if there is another pole at $\Theta =(2/3 -1/H)\pi i$ which meets its crossed
partner at the top of the diagram but is separated from it at the bottom
of the diagram. However, moving down the diagram, the coefficient of this pole
has the wrong sign to be interpreted as a bound state until it is crossed by
a zero represented by a dotted line. There the coefficient changes sign
and remains positive up to the bottom of the diagram.
A reasonable interpretation of this is that near the $d_4^{(3)}$ theory
this pole looks like the one appropriate for the $12\rightarrow 2$ coupling
but, far away it does not. A reasonable hypothesis amends the bootstrap
principle to include just those poles which never change sign over the
 whole interval. At first sight this seems strange. However, changing $\beta$
is actually equivalent to adjusting $\hbar$ (remember,
there is no classical coupling really), and therefore in a sense it is merely
being suggested that the structure of the quantum field theory should
be independent of a particular scale choice for $\hbar$. It would be
difficult to check this statement in perturbative field theory
because the zero in a pole coefficient is hard to find. Neverthess,
Delius, Grisaru and Zanon do give preliminary perturbative arguments
for the \lq floating' masses \refs{\rDGZa}.
On the other hand, there is another argument,
based on the Coleman-Thun mechanism
\NRF\rCTa\CTa\refs{\rCTa}, which suggests that a pole with
an indefinite coefficient might be best thought of
as a  double pole with a compensating zero. There is no time to pursue
this argument here but it is described in some detail in ref\refs{\rCDSa}.

The bootstrap principle applied to the coupling $11\rightarrow 2$ also
yields the third S-matrix element $ S_{22}(\Theta  )$ whose diagram is

\epsffile{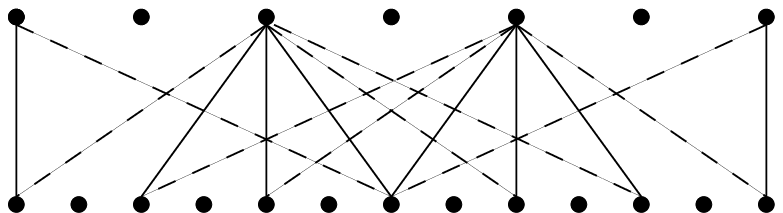}
\centerline{$S_{22}(\Theta  )$}

\noindent with the corresponding algebraic form
\eqn\gtwosthree{\eqalign{S_{22}(\Theta  )&={(0)\ (2H/3-2)\over (H-6)(4-H/3)}
\, {(2)\ (2H/3) \over (H-4)(6-H/3)}\, \cr
&\qquad \times \ {(H/3)\ (H-2)\over (4H/3-6) (4)}\, {(H/3+2)\ (H)\over
(H/3-4)(6)}\cr}}
This matrix element is quite fascinating. There are a number of poles but all
but two of them have coefficients which change sign. The poles  (at $\Theta
=2\pi i/3$, corresponding to the self-coupling $22\rightarrow 2$) which
do not change sign fail to do so because two zeroes happen to collide.
 All the other poles can be accommodated within the extended
Coleman-Thun scheme.

The other non simply-laced cases have all been listed elsewhere
\refs{\rFKMa , \rDGZa ,\rCDSa} and
will not be mentioned explicitly here. In every case, the generalised bootstrap
principle alluded to above is consistent and all poles not included in the
bootstrap have a plausible explanation within the extended Coleman-Thun scheme.

One intriguing question in all of this is, what (if anything) replaces the
beautiful structure surrounding the Coxeter element which  plays such
a unifying r\^ole in the simply-laced cases?
What replaces the formula \doreys ? Presumably, whatever the structure is,
it transcends the root lattices of the pair but offers a geometrical
setting for both of them (see \refs{\rDf}).

Finally, it is worth mentioning that there
is a convenient way to parametrise the coupling angles, and
the floating masses
\NRF\rDe\De\refs{\rDe}.
Define first, generalising the earlier notation,
\eqn\Bgeneral{B_g(\beta )\equiv B(\beta ; g,  g^{\vee})={2\beta^2\over
\beta^2 +4\pi (h/ h^{\vee})},}
where $g$ and $g^{\vee}$ have extended Dynkin diagrams related by
\duality , and associated Coxeter numbers $h$ and $h^{\vee}$, respectively.
Then, the following identity is true
\eqn\Bdual{B(\beta ; g, g^{\vee}) =2-B(4\pi /\beta ; g^{\vee}, g).}
The duality of the coupling angles for the positive definite poles
is then rendered transparent by
setting
\eqn\Bcoupling{\Theta _{ab}^c(\beta )={2-B_g\over 2}\Theta _{ab}^c(0)+
{B_g\over 2}\Theta _{ab}^c(\infty).}
A straightforward comparison with the coupling angles for the
$g_2^{(1)}\, -\, {d_4^{(3)}}$ case yields the consistent choice
$${1\over H(\beta )}={1\over 12}
\left(2-{B_{g_2^{(1)}}(\beta )\over 2}\right).$$

\newsec{A word on solitons}

If complex solutions to the affine Toda field equations are permitted
then there is a whole extra dimension to the Toda activity. At
first sight, the idea of allowing the Toda field to be complex
is unattractive since the classical hamiltonian will not be positive
definite and it is not immediately clear how such solutions
ought to be interpreted, or what their r\^ole in the quantum
Toda theory might be. On the other hand, it has been pointed out by
Hollowood
\NRF\rHa\Ha\refs{\rHa} that the soliton solutions, although complex, actually
have real energy  and momentum associated with them, despite the
fact that their energy-momentum density is complex. In addition,
the masses associated with the solitons are closely related to
the particle masses in the real theory and their couplings, in the
sense of a fusing rule, are also identical to the  couplings of
the real particles, at least for the $ade$ sequence of possibilities.
The static solitons are labelled by \lq topological charges' corresponding
to weights of the fundamental representations of the Lie algebra
underlying the Toda theory. This is quite easy to check but there is
something of a mystery associated with the topological charges in
the sense that complete sets of weights are only rarely found
(in the $a_n^{(1)}$ sequence of theories, and even then only in the
smallest dimension reresentations). This puzzle will be
mentioned again at the end of this section.

To see that the possibility of soliton solutions exists is not
difficult. It is enough to note that the Toda potential has local
stationary points whenever the field is constant and taken
to be
\eqn\todavac{\phi = {2 i\pi \lambda \over \beta}\qquad \hbox{with}\
\lambda\cdot\alpha_k=\hbox{integer}, \ k=0,1,....r\ .}
At each of these values of the field, the potential vanishes.
Soliton solutions to the equations of motion interpolate pairs
of these \lq vacua'  with, typically,
$$\phi (-\infty ,t)=0\qquad \phi (\infty ,t)={2 i\pi \lambda \over \beta .}$$
The topological charge is defined to be
\eqn\topcharge{\lambda = {\beta\over 2 i\pi}\left(\phi (\infty ,t)-
\phi (-\infty ,t)\right).}
At each of these values of the field, the potential vanishes.
Notice that this set up generalises the situation to be found
in the sine-Gordon theory which may be regarded as a purely imaginary
version of the $a_1^{(1)}$ affine Toda theory. Note, too, that
the sine-Gordon theory supplies the only example for which all the soliton
solutions are effectively real.
\bigskip
Each of the affine Toda theories contains soliton solutions and many
of the solutions have been catalogued elsewhere
\NRF\rMMa{\ACFGZ\semi\MMa\semi\CZa}\refs{\rMMa}. For definiteness and ease of
computation, the $a_n^{(1)}$ types will be illustrated here using
the so-called Hirota method as it was originally adopted by Hollowood. This
relies on the ansatz
\eqn\hirota{\phi = -{1\over  \beta}\sum_0^r \, \alpha_k \ln \, \tau_k}
for which the (time-independent) Toda field equations reduce to:
\eqn\hirotaeqn{\sum_0^r\alpha_k\left({\tau_k^{\prime\prime}\over \tau_k}
-{\tau_k^{\prime}\tau_k^{\prime}\over \tau_k^2}+\prod_l \tau_l^{-
\alpha_k\cdot\alpha_l}\right)=0\ .}
These are the relevant equations for static solitons. Using the
explicit form of the $a_r^{(1)}$ Cartan matrix \hirotaeqn\ may be rewritten
$$\sum_0^r\alpha_k\left({\tau_k^{\prime\prime}\over \tau_k}
-{\tau_k^{\prime}\tau_k^{\prime}\over \tau_k^2}+{\tau_{k-1}\tau_{k+1}
\over \tau_k^2}\right) =0\ ,$$
and solved by setting
\eqn\hirotachoice{{\tau_k^{\prime\prime}\over \tau_k}
-{\tau_k^{\prime}\tau_k^{\prime}\over \tau_k^2}+{\tau_{k-1}\tau_{k+1}
\over \tau_k^2}=1\quad\hbox{for}\ k=0,1,2\dots ,r\ ,}
where
$$\tau_k=1+\Omega_k e^{\sigma x +x_0},$$
provided
\eqn\recurrence{\eqalign{\sigma^2 \Omega_k -2\Omega_k +\Omega_{k-1}+
\Omega_{k+1}
&=0\cr
\Omega_{k-1}\Omega_{k+1}-\Omega_k^2&=0\cr
\Omega_{k+r+1}&=\Omega_{k}.\cr}}
The last pair of eqs\recurrence\  are solved by taking
$$\Omega_k^{(a)}=e^{2i\pi a k/r+1}=\omega^{ak}\qquad \hbox{for
each choice}\ a= 1,2,
\dots ,r\ ,$$
where $\omega$ is the primitive $r+1$st root of unity.
The first of the eqs\recurrence\  then imply a corresponding constraint
on $\sigma$ leading (for each choice of $a$) to
\eqn\sigmavalue{\sigma^{(a)}=2\sin{\pi a\over r+1}.}
The replacement $x\rightarrow -x$ gives another solution, and $x_0$ is
an arbitrary constant.
Assembling all these pieces, there is a solution for each $a$ of the
form \hirota :
\eqn\asoliton{\phi^{(a)} = -{1\over  \beta}\sum_0^r \, \alpha_k \ln
(1+\omega^{ak}e^{\sigma^{(a)}x+x_0^{(a)}})=-{1\over
\beta}\sum_1^r \, \alpha_k \ln
\left({1+\omega^{ak}e^{\sigma^{(a)}x+x_0^{(a)}}
\over 1+e^{\sigma^{(a)}x+x_0^{(a)}}}\right).}
These solutions are generally complex. The same solutions may be
obtained via a more sophisticated and general algebraic
method given by Olive, Turok and
Underwood based on the work of Leznov and Saveliev
 who
pioneered a  general approach to Toda wave equations some years ago
\NRF\rLSa\LSa
\NRF\rOTUa\OTUa
\NRF\rOTUb\OTUb\refs{\rLSa -\rOTUb}.
Moreover, it appears there are no other single soliton solutions to
be found using the more general techniques. This is perhaps surprising
given the special nature of the ansatz \hirota\  and the particular choice
of solution within it, eq\hirotachoice .
\bigskip
To calculate the energy of these solutions, it is extremely
convenient to use a formula for the energy-momentum tensor
established in the article by Olive, Turok and Underwood
\refs{\rOTUa}, using arguments
rooted in conformal Toda field theory. They found
\eqn\em{T_{\mu\nu}=(\eta_{\mu\nu}\partial^2 -\partial_{\mu}
\partial_{\nu})C,}
where the function $C$ for solitons is given by
$$C={-2\over \beta^2}\sum_0^r\ln\tau_k.$$
Using this, the energy of a static soliton (ie its mass)
can be calculated
\eqn\solitonmass{M^{(a)}=\int_{-\infty}^\infty dx\, T_{00}={\partial
 C\over \partial x}
\biggm|_{-\infty}^\infty ={2\over\beta^2}\sum_0^r {\tau_k^\prime
\over \tau_k}
\biggm|_{-\infty}^\infty = {2(r+1)\over \beta^2}\sigma^{(a)}.}
It is worthy of note that the mass is real despite the fact
that the energy density is complex and, moreover, each mass is proportional to
the mass of a corresponding elementary scalar particle in the real coupling
Toda theory provided the label $a$ is suitably interpreted.
\bigskip
To provide the
interpretation, first recall from lecture (1)
that the scalar particles of the real coupling
Toda theory are associated with
the fundamental representations of the Lie algebra $a_n$, the lightest
particles corresponding to  the smallest ($n+1$ dimensional) representation
of the algebra or its conjugate, the next lightest to the representations
of dimension $n(n+1)/2$, and so on. The solitons, on the other hand,
are labelled naturally by their topological charges which may
be calculated from the explicit solutions using \topcharge . The calculation
of the topological charges is slightly tricky and must be performed with
some care. First of all note that the argument of the logarithm in eq\asoliton\
must never vanish or diverge for any choice of $x$, otherwise the solution will
be singular. This requires that the constant $x_0^{(a)}$ (which may be
complex) has an imaginary part which is not entirely arbitrary; it is confined
to regions in the range $[0,2\pi ]$ whose boundaries correspond to
those choices of Im$x_0$ for which at least one of the logarithmic
arguments will vanish or diverge. Hence, the number of such boundary
points provides the maximum number of possibly different topological charges
which might be described by the solution \asoliton . Provided the boundary
points are avoided, the arguments of the logarithms change continuously,
the logarithm cannot jump its branch and the topological charge is
defined uniquely. McGhee
\NRF\rMc\Mc\refs{\rMc} has
calculated all the topological charges that are possible given \asoliton ,
and has confirmed that the topological charges of the solution whose mass
corresponds to that of the classical particle $a$ do indeed lie among
the weights of the associated representation. However, he has also noted
that the total number of possible topological charges typically
falls far short of the number required to fill up the whole representation.
Indeed there is a neat formula for the total number of topological charges
obtainable:
$$\hbox{number\ of\ charges\ of\ type\ }a\ =\ {n+1\over\hbox{gcd}(a,n+1)}.$$
Indeed, the only representation with its full complement of topological
weights is the representation of dimension $n+1$, or its conjugate.
\bigskip
McGhee has also examined a number of other theories and has found that
the Hirota solution in all other cases always fails to fill the associated
weight set and often the discrepancy is huge
\NRF\rMg\Mg\refs{\rMg}. A selection of the results
are given below, the numbers below the Dynkin diagram points denoting the
number
of possible topological charges obtainable via the Hirota ansatz.

$$\dddiv 4.2.4.4. \qquad\qquad\ddei 9.4.6.2.6.9.$$
$$\ddeii 10.8.2.2.4.6.14.$$
$$\ddeiii 18.6.8.2.6.10.14.26.$$

For the $e_8$ case the topological charges present are a small fraction of the
conjectured total. If it is really the case that the approach of Olive,
Turok and Underwood
captures all the solitons but is effectively equivalent to the Hirota ansatz,
then one must take seriously the \lq gaps' in the topological spectrum.
\bigskip
It has been suggested that there might be a consistent quantum field theory
corresponding to the complex Toda theories (a generalisation of the sine-Gordon
situation), in which the particle spectrum consists of multiplets corresponding
to the  representations of a quantised affine Lie algebra
(as is the case for the sine-Gordon theory in which the soliton and
anti-soliton
are a doublet of $su_q(2)$). This would appear to be natural but at the same
time
very mysterious without a detailed mechanism to explain
the
enormous gaps in the classical soliton spectrum.
Presumably, the classical spectrum must be enhanced in the quantum theory.
Hollowood
\NRF\rHb\Hb\refs{\rHb}, and Bernard and LeClair
\NRF\rBLa{\BLa\semi\BLb}\refs{\rBLa} have presented arguments suggesting that
the quantum theory ought to have such an enlarged spectrum. Certainly, it
is possible to obtain solutions to the Yang-Baxter equations,
based on quantum group ideas. There is no time
to consider these arguments here but one ought to bear in mind the intriguing
behaviour of the real coupling non simply-laced theories which would appear
to be difficult to mirror in the complex theory, since it is not at all clear
which quantum group one ought to be choosing. There is evidence that the
quantum
corrections to the soliton masses preserve the classical mass ratios for the
$ade$ cases
but fail to do so for the others. Hollowood
\NRF\rHf\Hf\refs{\rHf}, using old ideas of Dashen, Hasslacher and
Neveu, has calculated the
lowest order quantum mass corrections to
the soliton mass spectrum for $a_n^{(1)}$. Remarkably,
the mass corrections do not spoil the mass ratios.
Very recently, Mackay and Watts, and Delius
and Grisaru
\NRF\rMWa{\Wd\semi\MWa\semi\DGa}\refs{\rMWa}
 have performed similar calculations for non simply-laced
solitons and the classical mass ratios are not preserved.
One would imagine
that something akin to the duality going on in the real coupling theories
should persist provided the quantum field theory of the complex theories
really makes sense. It is conceivable a truncated spectrum will be necessary
in order to permit the S-matrix elements to enjoy floating bound-state poles.
That some truncation of the spectrum might be needed
is also indicated by a need for unitarity in a theory with a non-hermitean
Hamiltonian; an apparently
serious fault which might be alleviated by removing parts of the spectrum
to leave a unitary core.

It remains to be seen how this extremely interesting story will unfold.

\newsec{Other matters}

There are several interesting developments which cannot be described here.
For example, a full understanding of the quantum field theory would
require much more than the S-matrix/conserved quantity considerations
presented here. Indeed, there is a sizeable literature concerning the
calculation of form factors for Toda theory and related topics (for example,
see
\NRF\rSg{\Sg\semi\FMSa\semi\DMa\semi\Kd
\semi\Sh}\refs{\rSg}).

What happens if affine Toda field theory is restricted to a segment
of the real line, or to a half-line? The general question of integrability
in the presence of boundary conditions has its own literature but some recent
articles dealing specifically with Toda theories are given in
\NRF\rGZa{\GZa\semi\Gd\semi\FKc\semi\FKd\semi\Sk\semi\CDRSa\semi\CDRa}
refs\refs{\rGZa}. Surprisingly, there are strong constraints on the
possible form of the boundary condition maintaining classical integrability,
but it is not clear how these will affect the quantum theory---another
question to be resolved in the future.

\newsec{Acknowledgements}

I am grateful to the organisers of the school for the opportunity
to talk about Toda field theories, and to many colleagues and students
for stimulating interactions. In particular, I
would like to thank Harry Braden, Patrick Dorey, Richard Hall,
Tim Hollowood, Niall Mackay, William McGhee, Rachel Rietdijk, Ryu Sasaki,
G\'erard Watts, and Robert Weston for enjoyable discussions
and collaborations. I am
indebted to Patrick Dorey for various pictorial representations
of the singularities of S-matrix elements some of which have
been used in these lectures to illustrate the dual pairs of non simply-laced
models.

\listrefs
\end